\documentclass[a4paper,11pt]{article}
\usepackage{jcappub}
\usepackage{amssymb,amsmath,amsthm,amsfonts,mathrsfs}
\numberwithin{equation}{section}
\usepackage{enumerate}
\usepackage{bbm}
\usepackage{tensor}
\newcommand{\be}{\begin{eqnarray}}
\newcommand{\ea}{\end{eqnarray}}
\newcommand{\ben}{\begin{equation*}}
\newcommand{\een}{\end{equation*}}
\newcommand{\bean}{\begin{eqnarray*}}
\newcommand{\eean}{\end{eqnarray*}}
\def\bal#1\eal{\begin{align}#1\end{align}}

\newcommand{\bsub}{\begin{subequations}}
\newcommand{\esub}{\end{subequations}}

\newcommand{\disfrac}[1][2]{\displaystyle\frac}

\newcommand{\non}{\nonumber}

\newcommand{\sech}{\textrm{sech}}
\newcommand{\p}{\partial}
\title{Conditional symmetries in axisymmetric quantum cosmologies with scalar fields and the fate of the classical singularities}

\author[a]{Adamantia Zampeli}
\author[a]{Theodoros Pailas}
\author[a]{Petros A. Terzis}
\author[a]{T. Christodoulakis}

\emailAdd{azampeli@phys.uoa.gr}
\emailAdd{teopailas879@hotmail.com}
\emailAdd{pterzis@phys.uoa.gr}
\emailAdd{tchris@phys.uoa.gr}

\affiliation[a]{Nuclear and Particle Physics Section, Physics
Department,\\
University of Athens,15771 Athens, Greece}

\abstract{In this paper, the classical and quantum solutions of some axisymmetric cosmologies coupled to a massless scalar field are studied in the context of minisuperspace approximation. In these models, the singular nature of the Lagrangians entails a search for possible conditional symmetries. These have been proven to be the simultaneous conformal symmetries of the supermetric and the superpotential. The quantization is performed by adopting the Dirac proposal for constrained systems, i.e. promoting the first-class constraints to operators annihilating the wave function. To further enrich the approach, we follow \cite{Christodoulakis:2012eg} and impose the operators related to the classical conditional symmetries on the wave function. These additional equations select particular solutions of the Wheeler-DeWitt equation. In order to gain some physical insight from the quantization of these cosmological systems, we perform a semiclassical analysis following the Bohmian approach to quantum theory. The generic result is that, in all but one model, one can find appropriate ranges of the parameters, so that the emerging semiclassical geometries are non-singular. An attempt for physical interpretation involves the study of the effective energy-momentum tensor which corresponds to an imperfect fluid.}

\begin{document}
\maketitle
\flushbottom
\section{Introduction}
\subsection{General considerations}
An important advance in physics was the formulation of general relativity, which successfully describes the classical gravitational field. A remarkable feature of this theory is that it predicts the existence of singularities, i.e. situations in which spacetime breaks down in the sense that some curvature scalars as well as other physical quantities diverge. This phenomenon has been proven by the various singularity theorems to be a quite general occurrence, not related to any particular assumption of symmetry \cite{Hawking:1973uf}. A manageable treatment of singularities usually appears in black holes and in cosmological spacetimes. The hope is that quantum effects will become important near the singularity regime and thus soften or even eliminate the problem. 

Quantum cosmology, which is a branch of quantum gravity dealing with cosmological models, is a simplified approach constructed by assuming a high degree of symmetry. This allows to freeze all but a finite number of degrees of freedom and therefore results in a manageable reparametrization invariant canonical quantum theory. Despite the considerable simplification, the crucial problem of unitarity is still present; since the Wheeler-DeWitt equation is of a Klein-Gordon type differential equation, there is no obvious way to define a positive-definite probability. A quite interesting and elegant investigation of this issue is given in \cite{BARVINSKY1993237}, where the solutions of the Wheeler-DeWitt equation are related to the wave function of the true physical variables; i.e. the ones remaining after an appropriate choice of a gauge fixing condition. This approach was put in action in \cite{PhysRevD.89.043526} and more recently in \cite{PhysRevD.93.083519}; the first work has an overlap with the present work in the case of FLRW cosmology. As we explicitly show in the appendix \ref{comparisonbarvinsky}, the wave functions we find are identical to the ones in \cite{PhysRevD.89.043526}, even though we take a different tributary from the same river. Barvinsky's approach in \cite{BARVINSKY1993237} is based in the reduced phase space quantization method, which is accomplished by a gauge choice and the use of path integrals. This allows for the transition from the physical degrees of freedom to the initial superspace variables and to the original Wheeler-DeWitt solution. On the other hand, in this work we do not make any such gauge choice. Instead, we use the symmetries of the supermetric in order to arrive at the final wave function.

A usual semiclassical treatment of the solutions of the Wheeler-DeWitt equation is by employing the WKB approximation and the construction of wave packets \cite{Kiefer:1988tr}. In this approximation, it was shown that the singularity can be avoided in homogeneous and isotropic spacetimes (see e.g. \cite{Kiefer:2010zz}). Another approach to the solution of the Wheeler-DeWitt equation is to follow Bohm's theory \cite{Bohm:1951xw,Bohm:1951xx,Bohm:1984aa} and construct deterministic quantum trajectories in the superspace. It was shown that the singularity can also be avoided, again for a homogeneous and isotropic metric coupled to scalar fields or filled with dust and radiation \cite{Vink:1990fm,Colistete:1997sf,AcaciodeBarros:1997gy}. Similar results are obtained from loop quantum cosmology for homogeneous isotropic models with a scalar field \cite{Bojowald:2001xe,Ashtekar:2006rx,Ashtekar:2006uz,Ashtekar:2006wn,Ashtekar:2007em,Ashtekar:2008zu,Bojowald:2015iga}.

In this work, the starting point are 4-dimensional spacetimes minimally coupled to a massless scalar field. We focus on the study of isotropic models as well as anisotropic. We adopt a minisuperspace approximation of gravity for quantum geometrodynamics. In these classical spacetimes, the lapse function is not gauge fixed. This allows for the presence of extra symmetries in the configuration space called conditional symmetries (see e.g. \cite{Kuchar:1982eb,Christodoulakis:2000xe,Christodoulakis:2001sc,Christodoulakis:2001um,Christodoulakis:2012eg,Christodoulakis:2013sya}). In \cite{Christodoulakis:2013xha}, the relation between the Lie point symmetries and the conditional symmetries of the minisuperspace was established which in the constant potential lapse parametrization coincide with the conditional symmetries in the phase space. A different approach regarding the Lie point symmetries and the gauge fixing of the lapse function was taken in \cite{Paliathanasis:2013tza,Terzis:2015mua} and applied in \cite{Paliathanasis:2015gga,Paliathanasis:2015arj,Tsamparlis:2013aza}.
The system of the Einstein's field equations is solved by using the first integrals of motion associated to the conditional symmetries; this method facilitates the solution of these equations. For the quantization of the system, we first promote the first-class constraints to operators annihilating the wave function, according to Dirac \cite{Dirac:113811}. Then, the generators of the conditional symmetries are also promoted to operators, thus providing a system of quantum constraints with additional eigenvalue equations on the wave function. The outcome is a unique wave function, not containing arbitrary functions, which is used to find a semiclassical spacetime. This is done by writing it in polar form and setting up the corresponding semiclassical equations following Bohm's approach to quantum theory. This semiclassical solution is studied with respect to two aspects: firstly, we check whether the singularity can be avoided. Secondly, the effective energy-momentum tensor is interpreted as an energy-momentum tensor of an imperfect fluid. Therefore, even if we started with a classical spacetime filled with a massless scalar field, the semiclassical spacetime can acquire a different physical content.

\subsection{Classical treatment}
We assume a 4-dimensional cosmological spacetime with line element of the form
\begin{equation}\label{bianchi_metric}
ds^2 = -N^2 (t) dt^2 +  \gamma_{\alpha \beta}(t) \sigma^\alpha_i(x) \sigma^\beta_j(x) dx^i dx^j, \ i,j = 1,2,3
\end{equation}
where $N(t)$ is the lapse function, $\gamma_{\alpha \beta}(t)$ is a positive-definite diagonal $3 \times 3$ axisymmetric matrix of the dependent dynamical variables and $\sigma_i^\alpha (x)$ are the invariant one-forms which satisfy the relation 
\begin{equation}\label{sigma_relation}
\sigma^\alpha_{i,j} -\sigma^\alpha_{j,i} = C^\alpha_{\beta \gamma} \sigma^\beta_j \sigma^\gamma_j.
\end{equation}
For each Bianchi type the one-forms $\sigma^\alpha$ are well-known and can be found e.g. in \cite{Ryan:1975jw}. The general classical solution of most Bianchi types in vacuum has been given in \cite{Terzis2006cart,Christodoulakis:2006vi,Terzis:2008ev,Terzis:2010dk}. The action integral and the corresponding equations of motion are given by\footnote{We are using units $8 \pi G=\kappa^4$.}
\begin{equation}\label{total_action}
S_{tot} = S_{grav} +S_{mat}=\int d^4 x \sqrt{-g} \left(R-\frac{1}{2} g^{\mu\nu} \partial_\mu \phi \partial_\nu \phi \right), \quad 
E_{\mu\nu} = -\frac{1}{4} g_{\mu \nu}g^{\kappa \lambda} \partial_\kappa \phi \partial_\lambda \phi +\frac{1}{2} \partial_\mu \phi \partial_\nu \phi,
\end{equation}
where $R$ is the Ricci scalar, $E_{\mu \nu}$ the Einstein tensor and we have assumed the existence of a massless scalar field $\phi$. The variation of the action with respect to $\phi$ results to the well-known equation of motion for the scalar field minimally coupled to gravity, the Klein-Gordon equation, $g^{\mu \nu} \nabla_\mu \nabla_\nu \phi=0$. Integrating the spatial coordinates $x^i$, the action reduces to $S_{tot}= \int dt\ L$, where $L$ is a Lagrangian of the form
\begin{equation}\label{singular_lagrangian}
L=\frac{1}{2 N} G_{\alpha \beta} (q) \dot{q}^\alpha \dot{q}^\beta - N V(q),
\end{equation}
and $G_{\alpha \beta} (q)$ is the supermetric defined on the configuration space of the dependent variables $q^\alpha (t)$. Of course, there is always the question of whether this reduced Lagrangian is valid: one must check that the reduction of the Einstein plus matter equation is equivalent to the Euler-Lagrange equations. This equivalence has been checked for all models we present in this paper. The Lagrangian \eqref{singular_lagrangian} is time-reparametrization invariant, i.e. retains its form under the transformation $t = f(\tilde{t})$, if one also changes the lapse function and the dependent variables as
\begin{equation}
N(t)\rightarrow \tilde{N} (\tilde{t}) = N (f(\tilde{t})) f' (\tilde{t}), \quad q^\alpha (t) \rightarrow \tilde{q}^\alpha(\tilde{t}) =q^\alpha (f(\tilde{t})).
\end{equation}
These can be considered as a type of gauge transformations owing their existence to the presence of constraints in the system. The primary constraint is the conjugate momentum of the lapse function
\begin{equation}
p_N =\frac{\partial L }{\partial \dot{N}} \approx 0,
\end{equation}
and the demand for its conservation in time leads to the secondary constraint $\mathcal{H}$ 
\begin{equation}
\dot{p}_N = \{p_N,H \} \Rightarrow \mathcal{H} =\frac{1}{2} G^{\alpha \beta} (q) p_\alpha p_\beta  + V(q) \approx 0,
\end{equation}
known as the Hamiltonian constraint because of the relation
\begin{equation}
H = p_\alpha \dot{q}^\alpha - L = N \left( \frac{1}{2} G^{\alpha \beta} (q) p_\alpha p_\beta  + V(q)\right) \equiv N \mathcal{H}.
\end{equation}
The Dirac procedure is then terminated, because the demand for conservation in time of $\mathcal{H}$ does not give any new constraints. Therefore, the theory has two constraints, one linear and one quadratic in momenta and they are first class due to the relation $\{ p_N, \mathcal{H}\} \approx 0$.
In the context of singular systems, it has been shown that the variational symmetries of the action \eqref{singular_lagrangian}, as well as the Lie point symmetries of the Euler-Lagrange equations, result in the appearance of conditional symmetries; these are defined as the simultaneous symmetries of the supermetric and the superpotential with the same conformal factor
\begin{equation}\label{conditional_symmetries}
\mathcal{L}_\xi G^{\alpha \beta} (q) = \tau (q) G^{\alpha \beta} (q), \quad
\mathcal{L}_\xi V (q) = \tau (q) V(q),
\end{equation}
where $\mathcal{L}_\xi$ is the Lie derivative operator on the configuration space. The symmetry generators, say $\xi_i$ ($i$ numbers the different $\xi_i$'s) satisfy a Lie algebra of the form
\begin{equation}
[ \xi_i , \xi_j ] = c^k_{ij} \xi_k,
\end{equation}
where $c^k_{ij}$ are the structure constants. To each one of the generators, there corresponds a conserved quantity $Q_i = \xi_i^\alpha p_\alpha$ linear in the conjugate momenta $p_\alpha = \frac{\partial L}{\partial \dot{q}^\alpha}$; the corresponding Poisson bracket satsify the same algebra as their generators. Exploiting the freedom given by the time parametrization invariance, one can choose a new lapse $n=N V$ which implies $\bar{G}_{\alpha \beta}=V G_{\alpha \beta}$, $\bar{V} = 1$. The new scaled Lagrangian reads 
\begin{equation}
L=\frac{1}{2 n} \bar{G}_{\alpha \beta} (q) \dot{q}^\alpha \dot{q}^\beta - n,
\end{equation}
and now the generators $\xi^\alpha$ retain their form and become Killing fields of the scaled supermetric
\begin{equation}
\mathcal{L}_\xi \bar{G}^{\alpha \beta} (q) =0.
\end{equation}
If the action corresponds to a gravitational minisuperspace model, then there is an extra symmetry (the one corresponding to the constant scaling of all scale factors), reflected in the existence of a homothetic vector field. Note that this symmetry already known to exist at the level of the equations of motion. The new charges $Q_i$ of the scaled Lagrangian still satisfy the same relations as well as the same Poisson bracket algebra. From the evolution of the scalars $Q_i$,
\begin{equation}
\frac{d Q_i}{dt} = \{Q_i , H \} = -\omega n,
\end{equation}
where $\omega=0$ for the Killing vector fields and $\omega=1$ for the homothetic, we get for each Killing vector field an integral of motion
\begin{equation}\label{integralofmotion}
Q_i =\kappa_i,
\end{equation}
while for the homothetic Killing feld a rheonomic integral of motion
\begin{equation}\label{rheonomicintegral}
Q_h = \kappa_h - \int n(t) dt,
\end{equation}
as it was shown in \cite{Christodoulakis:2012eg,Christodoulakis:2013xha}. The advantage of solving eqs \eqref{integralofmotion}, \eqref{rheonomicintegral} is that the system of the dependent variables is of first order in the velocities/momenta and it is much more easily solved. Furthermore, if there exist enough integrals of motion, the problem of determining $q^i (t)$ reduces to an algebraic one \cite{Terzis:2014cra}.

Finally, one can also find higher order generators, i.e. irreducible tensor Killing fields leading to new conserved quantities that do not reduce to the first order ones \cite{Terzis:2015mua}. An important quantity is the Casimir invariant which commutes with all the $Q_i$'s
\begin{equation}
\{ Q_i, Q_{cas}\} =0,
\end{equation}
and is constructed by an element of the universal enveloping algebra spanned by $\xi_i$'s. 
 
In summary, for the classical analysis we first find the Lagrangian of the full gravitational system coupled to the massless scalar field in order to rescale it and obtain a Lagrangian of constant superpotential. The Killing and homothetic symmetries of the resulting minisuperspace are studied and the corresponding conserved quantities are constructed and used to solve for the configuration space variables. The final solution is expressed in the initial Lagrangian variables and replaced in the spacetime metric. The resulting spacetime is then studied for singularities.
\subsection{Quantization method}
The quantization of the system is performed in the new scaled Lagrangian and the corresponding Hamiltonian by adopting Dirac's procedure for constrained systems. This means that the first-class constraints of the theory are imposed as conditions on the wave function and lead to the equations
\begin{align}
&\hat{p}_N \Psi(q,N) \equiv - i \frac{\partial}{\partial N} \Psi (q,N) = 0 \Rightarrow \Psi \equiv \Psi (q)  \label{momentum_constraint}, \\
&\hat{\mathcal{H}} \Psi (q)  \equiv \left(-\frac{1}{2} \Box_c^2+ 1 \right) \Psi (q) =0 .\label{hamiltonian_constraint}
\end{align}
Eq. \eqref{momentum_constraint} is known as the momentum constraint and leads to the independence of the wave function from the lapse function, while \eqref{hamiltonian_constraint} is the Wheeler-DeWitt equation and gives the dynamics of the system. In the above relations $\Box_c$ is the conformal Laplacian defined as
\begin{align}
\Box_c^2 \equiv \Box^2 + \frac{d-2}{2 (d-1)} R,
\end{align}
where $R$ is the Ricci scalar of the superspace. The choice for the factor ordering for the kinetic part of the Hamiltonian implied by the conformal Laplacian stems from the demands of (i) general covariance of the Lagrangian, (ii) hermiticity of the operators under the inner product of the form $\int d^n q \mu \psi^*_1 \psi_2$, with $\mu$ a proper measure and (iii) conformal invariance of the action. These conditions, as well as the demand that the momentum operators act on the far right, fix the form of the measure to be $\mu = \sqrt{|\det G_{\alpha \beta}|}$ \cite{Christodoulakis:1984gp,Christodoulakis:2001um,Christodoulakis:2012eg}. 

The critical step in the approach we follow is to promote the generators of the conditional symmetries \cite{Christodoulakis:2012eg} to operators, in addition to the constraints of the classical system, thus forming the following eigenvalue problem 
\begin{equation}\label{eigenvalueeqs}
\hat{Q}_i \Psi (q) \equiv -\frac{i}{2 \mu} (\mu \xi_i^\alpha \partial_\alpha +\partial_\alpha \mu \xi_i^\alpha) \Psi (q)= \kappa_i \Psi (q).
\end{equation}
For consistency with the classical system, the eigenvalues must be equal to the corresponding classical charges $\kappa_i$. Consequently, the classical algebra of the generators 
\begin{equation}
\{ Q_i ,Q_j\} = c^k_{ij} Q_k,
\end{equation}
becomes an isomorphic quantum algebra with the same structure constants. However, this quantum algebra through the consistency conditions $[\hat{Q}_i , \hat{Q}_j ] \Psi =(\kappa_i \kappa_j - \kappa_j \kappa_i)\Psi = 0$ leads to the integrability conditions \cite{Christodoulakis:2012eg}:
\begin{equation}\label{selection_rule}
c_{ij}^k \kappa_k =0,
\end{equation}
which works as a selection rule for the possible admissible subalgebras, since not all of the quantum generators can be imposed simultaneously on the wave function. 

The quantum system is formed by eqs \eqref{momentum_constraint}, \eqref{hamiltonian_constraint} and this set of the eigenvalue equations \eqref{eigenvalueeqs} of the operators belonging to the corresponding subalgebra. We note that the generator of the homothecy of the system is not quantised (due to its explicit time-dependence) and thus not used for the solution of the quantum system, even though it is used to obtain the classical solution. However, it is of great use to promote the Casimir invariant to operator and include it as an additional eigenvalue equation of the same form as \eqref{eigenvalueeqs}, in the cases where the quantum equations are not enough to provide a unique solution. This is also true for any Killing tensor of higher order that commutes with the operators $\hat{Q}_i$ of the corresponding subalgebra, if of course it exists.

\subsection{Semiclassical approximation and physical results}\label{Bohmian semiclassical analysis}
We now wish to use the wave function to extract some physical insight. Our aim is twofold: (i) check whether the classical singularities persist after the quantization and (ii) understand the matter content of the emergent spacetime by comparing the effective energy-momentum tensor with the corresponding one of an imperfect fluid. 

To this end, we perform an analysis based on Bohm's interpretation of quantum theory \cite{Bohm:1951xw,Bohm:1951xx,Bohm:1984aa} following \cite{Christodoulakis:2013sya,Christodoulakis:2014wba}. Bohmian interpretation in the context of quantum cosmology has also been considered elsewhere but with a different methodology (see e.g. \cite{Falciano:2015fja,Pinto-Neto:2013toa,PintoNeto:2004uf,Struyve:2015cwa,Vink:1990fm} and references therein). The Bohmian approach to quantum theory is well suited for quantum cosmology, since it does not presupposes the existence of a classical domain as it is necessary for the Copenhagen interpretation for the measurement process to be defined. In addition, this interpretation results in the definition of deterministic trajectories in the space of geometries. This allows the comparison of the classical versus semiclassical trajectories through the corresponding geometric properties of each geometry.

Bohmian mechanics starts with the assumption that the wave function is written in the polar form and by inserting it in the Schrodinger equation one obtains, separating real and imaginary parts, the continuity equation for the amplitude as well as a modified Hamilton-Jacobi equation for the phase which differs from the classical Hamilton-Jacobi equation by a term of quantum origin known as quantum potential. This equation defines the particle's quantum trajectories analogously to the classical equation which defines the classical trajectories of the particle.

To apply this procedure in minisuperspace models, we start with the polar form 
\begin{equation}\label{polar_form}
\Psi (q) = \Omega (q) e^{iS(q)},
\end{equation}
where $\Omega (q)$ is the amplitude, $S(q)$ is the phase of the wave function and $q^\alpha$ are the variables of the configuration space, and replace it to the Wheeler-DeWitt equation. The result is a modified Hamilton-Jacobi equation
\begin{equation}\label{modified_HJ} 
\frac{1}{2} G^{\alpha \beta} \partial_\alpha S \partial_\beta S -\frac{1}{2} \frac{\Box \Omega}{\Omega} +V =0,
\end{equation}
and a second equation that in ordinary quantum theory it is interpreted as the continuity equation 
\begin{equation}
G^{\alpha \beta} \partial_\alpha S \partial_\beta \Omega + \frac{\Omega}{ 2 \mu} \partial_\alpha (\mu G^{\alpha \beta} \partial_\beta S)=0.
\end{equation}
However, in this case probabilities are not defined therefore this equation cannot have the meaning of a continuity equation. 

Following Bohm one can observe that the quantity $\partial_\alpha S$ can be identified with the momentum of the system, thus assuming that the classical definition $p_\alpha = \partial_\alpha S$ is still valid. We can also recognise the second term as a new potential term coming from quantum effects known as quantum potential,
\begin{equation}
\mathcal{Q} (q) \equiv -\frac{1}{2 \Omega} \Box \Omega =- \frac{1}{2 \mu} \partial_\alpha (\mu G^{\alpha \beta} \partial_\beta)\Omega .
\end{equation}
This line of thinking can lead to the equations \cite{Christodoulakis:2013sya}
\begin{equation}\label{semiclassical_eqs}
\frac{\partial S}{\partial q^\alpha}=\frac{\partial L}{\partial \dot{q}^\alpha}.
\end{equation}
When the quantum potential vanishes (see section V of \cite{Christodoulakis:2013sya} for a full explanation), the solution of this set of equations coincides with the classical one, while for a non-vanishing quantum potential this is not true anymore.

In this work, the wave function \eqref{polar_form} is the solution obtained from each subalgebra. This is either already in polar form or it is written in this way in some approximation limits; thus the phase function and the amplitude are immediately read from $\Psi$. As it is expected, for those models for which $\mathcal{Q}=0$, the emerging semiclassical geometries are equivalent to the classical ones, while for non-zero potential we obtain a different solution. The new semiclassical spacetime can be studied for the fate of the classical singularities, which is a main issue in the study of the modern quantum cosmology. The other aspect of the semiclassical spacetime we then focus on is its matter content. To this end, we simulate the effective energy-momentum with that one of an imperfect fluid. 

In the remainder of the paper, the classical, quantum and semiclassical behaviour of the following models coupled to a massless scalar field is examined: the Kantowski-Sachs spacetime \ref{Massless scalar field in Kantowski-Sachs spacetime}, the FLRW geometry \ref{Massless field in the closed FLRW universe} as well as the Bianchi type II \ref{bianchi_II}, type V \ref{bianchi_V} and type VI \ref{bianchi_VI}
spacetimes. In section \ref{Physical Interpretation}, we focus on the study of the matter content of the semiclassical spacetimes obtained by the subalgebras with non zero quantum potential. Finally in section \ref{conclusions} the results are discussed and some conclusions are drawn based on the physical results. 

\section{Massless scalar field in Kantowski-Sachs spacetime}\label{Massless scalar field in Kantowski-Sachs spacetime}
\subsection{Classical treatment}
We first study the Kantowski-Sachs spacetime coupled to a massless scalar field. Its metric line element assumes the form
\begin{equation}
ds^2 = -N^2(t) dt^2 + a^2 (t) dr^2 + b^2(t) (d\theta^2 + \sin^2 \theta d\varphi^2),
\end{equation}
where $a(t), b(t)$ are scale factors depending only on time. Even though this metric has the form \eqref{bianchi_metric}, it does not belong to the Bianchi classification because the action of the group of isometries on the 3-dimensional spacelike surface is not simply transitive as it is necessary for a model to be characterized as Bianchi. This is reflected in the non-constancy of the structure functions appearing in \eqref{sigma_relation}, $C^3_{23} = \frac{\cot \theta}{2}$, with $\sigma^\alpha_i =diag (1,1,\sin \theta)$. 

Integrating over the spatial directions and discarding a term of total derivative the integrand of the total action \eqref{total_action} leads to the following Lagrangian
\begin{equation}
L=2 aN-\frac{4 b \dot{a} \dot{b}}{N} -\frac{2 a\dot{b}^2}{N} -\frac{ab^2 \dot{\phi}^2}{2 N}. 
\end{equation}
In the constant potential parametrization, the rescaled Lagrangian is written
\begin{equation}\label{constant_lag_kantowski}
L= -n + \frac{8 ab \dot{a} \dot{b}}{n} + \frac{4 a^2 \dot{b}^2}{n} + \frac{a^2 b^2 \dot{\phi}^2}{n},
\end{equation}
where we have set $n=-2 N a$. The Hamiltonian constraint is found to be
\begin{align}
\mathcal{H} = 1 - \frac{p_a^2}{16 b^2} + \frac{p_a p_b}{8a b} + \frac{p_\phi^2}{4 a^2 b^2} \approx 0,
\end{align}
the supermetric of the configuration space is read off the kinetic part of the Lagrangian as
\begin{align}
G_{\alpha \beta} =\left(
\begin{array}{cccc}
0 & 8ab &0\\
8ab & 8 a^2&0\\
0 & 0 & 2a^2 b^2\\
\end{array}
\right).
\end{align}
This superspace is conformally flat; the generators of the superspace symmetries become
\begin{equation}
\xi_1 =-a\partial_a+ b\partial_b, \quad
\xi_2 =-a\phi \partial_a + b\phi \partial_b-4 \ln a \partial_\phi, \quad
\xi_3 =\partial_\phi, \quad 
\xi_h =\frac{a}{2}\partial_a,
\end{equation}
where $\xi_i, \ i=1,2,3$ represent the Killing fields and $\xi_h$ the homothetic field. Their Lie bracket algebra is 
\begin{align}
[\xi_1,\xi_2]= 4 \xi_3, \quad
[\xi_2,\xi_3]= -\xi_1, \quad
[\xi_2,\xi_h]= 2 \xi_3,
\end{align}
and the system of the first integrals expressed in the velocity phase space variables is
\begin{align}
Q_1 &=\frac{8a b^2 \dot{a}}{n} = \kappa_1 \label{conserved_Q1}, \\
Q_2 &=\frac{8 a b^2 \left( \phi \dot{a} -a \dot{\phi} \ln a \right)}{n} =\kappa_2 \label{conserved_Q2},\\
Q_3 & = \frac{2 a^2 b^2 \dot{\phi}}{n}=\kappa_3, \label{conserved_Q3}, \\
Q_h &= \frac{4 a^2 b \dot{b}}{n}=\kappa_h - \int dt \ n(t),
\end{align}
where $\kappa_i, \ i=1,2,3,h$ are constants. The algebraic solution of the system gives us a relation between the variables $(a, \phi)$
\begin{equation}\label{condition_kantowski}
\phi = \frac{\kappa_2 + 4 \kappa_3 \ln a}{\kappa_1}.
\end{equation}
Inserting this relation in the initial Lagrangian \eqref{constant_lag_kantowski} results in a further reduced Lagrangian with two degrees of freedom instead of three:
\begin{align}
L_{red} = -n + \frac{16 \kappa_3^2 b^2 \dot{a}^2}{\kappa_1^2 n} + \frac{8 ab \dot{a} \dot{b}}{n} + \frac{4 a^2 \dot{b}^2}{n}.
\end{align}
This reduced Lagrangian is also valid in the sense that the emanating equations of motion are equivalent to those obtained from \eqref{constant_lag_kantowski} when the solution \eqref{condition_kantowski} is inserted. This is in general not the case for every Lagrangian and the equivalence has to be checked every time we assume a particular reduction . The advantage of using $L_{red}$ is that the problem becomes one with less degrees of freedom, thus easier to be solved. 

Following the same procedure, we find the Hamiltonian constraint corresponding to the reduced Lagrangian: 
\begin{align}
\mathcal{H}_{red} = 1- \frac{\kappa_1^2 p_a^2}{16 \lambda b^2} + \frac{\kappa_1^2 p_a p_b}{8 \lambda ab} - \frac{\kappa_3^2 p_b^2}{4 \lambda a^2} \approx 0,
\end{align}
where we have written for simplicity $\lambda^2=\kappa_1^2 -4 \kappa_3^2$ which is the Casimir operator of the algebra of the conserved quantities of Lagrangian \eqref{constant_lag_kantowski}. The corresponding supermetric is 
\begin{align}
G_{\alpha \beta} =\left(
\begin{array}{cccc}
\frac{32 b^2 \kappa_3^2}{\kappa_1^2} & 8ab \\
8ab & 8 a^2\\
\end{array}
\right),
\end{align}
and its symmetry generators are
\begin{align}
\zeta_1 &= -a \partial_a +  b \partial_b, \\
\zeta_2 & = -\frac{\kappa_1 \sinh (\frac{\lambda \ln a}{\kappa_1})}{b \lambda}\partial_a+ \frac{\cosh (\frac{\lambda \ln a}{\kappa_1}) + \frac{\kappa_1}{\lambda} \sinh (\frac{\lambda \ln a}{\kappa_1})}{a} \partial_b,\\
\zeta_3 & = -\frac{\kappa_1 \cosh (\frac{\lambda \ln a}{\kappa_1})}{b \lambda}\partial_a +  \frac{\sinh (\frac{\lambda \ln a}{\kappa_1}) + \frac{\kappa_1}{\lambda} \cosh (\frac{\lambda \ln a}{\kappa_1})}{a} \partial_b,\\
\zeta_h & = \frac{a}{2}\partial_a,
\end{align}
satisfying the following Lie algebra
\begin{align}
[\zeta_1,\zeta_2] = -\frac{\lambda}{\kappa_1} \zeta_3, \quad
[\zeta_1,\zeta_3] = - \frac{\lambda}{\kappa_1} \zeta_2, \quad
[\zeta_2,\zeta_h] = \frac{1}{2} \zeta_2 - \frac{\lambda}{2 \kappa_1} \zeta_3, \quad
[\zeta_3,\zeta_h] = - \frac{\lambda}{2 \kappa_1} \zeta_2 + \frac{1}{2} \zeta_3.
\end{align}
The solution of the system $Q_{red i}\equiv \zeta_i^\alpha p_\alpha=c_i, \ i=1,2,3,h $ is
\begin{align}
b& = \frac{c_1 \kappa_1}{\lambda a \left( c_3 \cosh \frac{\lambda \ln a}{\kappa_1} -c_2 \sinh \frac{\lambda \ln a}{\kappa_1} \right)} ,\\
\dot{b} & = \frac{c_1 \left( -c_3 \kappa_1^2 + 4 c_3 \kappa_3^2 + c_2 \kappa_1 \lambda + \left(c_2 \lambda^2 -c_3 \kappa_1\lambda \right) \coth \frac{\lambda \ln a}{\kappa_1} \sinh^{-1} \frac{\lambda \ln a}{\kappa_1} \dot{a} \right)}{\lambda^2 a^2 \left( c_2 -c_3 \coth \frac{\lambda \ln a}{\kappa_1} \right)^2},\\
n& = \frac{8 c_1 \dot{a}}{a \left( c_3 \cosh \frac{\lambda \ln a}{\kappa_1} -c_2 \sinh \frac{\lambda \ln a}{\kappa_1} \right)},\\
\int dt \ n(t)  & = c_h + \frac{c_1 \left(- c_3 \kappa_1 + c_2 \lambda + (c_2 \kappa_1-c_3 \lambda ) \coth \frac{\lambda \ln a}{\kappa_1}\right)}{2 \lambda \left( c_2 -c_3 \coth \frac{\lambda \ln a}{\kappa_1} \right)}.
\end{align}
plus a relation for the constants $c_3^2 - c_2^2 =16$ resulting from the demand that the solutions satisfy the constraint. This relation is the arithmetic value of the Casimir invariant of the algebra spanned by the generators of the reduced supermetric on shell. Now, by setting $c_2 =4  \sinh \kappa$ and selecting as a gauge condition $a= e^t$, the solution is found to be
\begin{align}
N =-\frac{c_1}{4 e^t \cosh^2 (\kappa - \frac{\lambda t}{\kappa_1})} ,\quad
b  =\frac{c_1 \kappa_1 }{4 \lambda e^t \cosh (\kappa - \frac{\lambda t}{\kappa_1})}, \quad
\phi  = \frac{\kappa_2 + 4 \kappa_3 t}{\kappa_1}.
\end{align}
This solution still contains redundant constants which are not necessary for the description of geometric features of the spacetime. Therefore, several changes of coordinates, allowed by the diffeomorphism covariance of the theory, are applied to absorb these constants. The final line element is written as
\begin{align}
ds^2 = -\frac{ \beta e^{\alpha T} }{\cosh^4 T} dT^2 + e^{-\alpha T} dr^2 +\frac{ e^{\alpha T} \beta}{\cosh^2 T}d\theta^2 + 
\frac{e^{\alpha T}}{\cosh^2 T} \sin^2 \theta d \varphi .
\end{align}
As it has been checked with the theorem of \cite{Papadopoulos:2005ft,Christodoulakis:2003wy}, this metric is characterized by two essential constants. Its Ricci scalar is 
\begin{align}
R =\frac{e^{-\alpha T} (\alpha^2 -4) \cosh^4 T}{2 \beta}.
\end{align}
The singularity of this spacetime appears for $T \rightarrow \infty$ for any value of the constants, apart from $\alpha^2=4$. We now proceed to the quantization and the semiclassical analysis of the results in order to find out whether the singularity can be removed.
\subsection{Canonical quantization and semiclassical analysis}
The quantization of the system is performed for the reduced Lagrangian. The admissible subalgebras which satisfy the relation \eqref{selection_rule} are the two-dimensional $(\hat{Q}_2, \hat{Q}_3)$ and the one-dimensionals $\hat{Q}_1,\hat{Q}_2,\hat{Q}_3$. However, the latter two are members of a higher dimensional admissible subalgebra and their results are not studied separately \cite{Christodoulakis:2012eg,Christodoulakis:2013sya}. We adopt this point of view for all the other models of the paper. Omitting the momentum constraint equation, since its impact is always the same on the wave function, the quantum equations on our disposal are
\begin{align}
\hat{Q}_1 \Psi &= i (-b \partial_b + a \partial_a) \Psi = \kappa_1 \Psi ,  \label{ksq1}\\
\hat{Q}_2 \Psi &= -\frac{i}{ab \lambda^2} \left(
b\left(\lambda^2 \cosh \frac{\lambda \ln a}{\kappa_1} + \kappa_1 \lambda \sinh\frac{\lambda \ln a}{\kappa_1}\right) \partial_b - a \kappa_1 \lambda \sinh\frac{\lambda \ln a}{\kappa_1} \partial_a\right) \Psi = \kappa_2 \Psi, \label{ksq2} \\
\hat{Q}_3 \Psi &= -\frac{i}{ab \lambda^2} \left(
b\left(\kappa_1 \lambda \cosh \frac{\lambda \ln a}{\kappa_1} +  \lambda^2 \sinh \frac{\lambda \ln a}{\kappa_1} \right) \partial_b - a \kappa_1 \lambda \cosh \frac{\lambda \ln a}{\kappa_1} \partial_a\right) \Psi = \kappa_3 \Psi, \label{ksq3} \\
\hat{\mathcal{H}} \Psi &=\left( -1 -\frac{\kappa_3^2}{4 a^2 b \lambda^2}\partial_b -\frac{\kappa_3}{4 a^2 \lambda^2} \partial_{bb} -\frac{\kappa_1^2}{16 a b^2 \lambda^2} \partial_a + \frac{\kappa_1^2}{8 ab \lambda^2} \partial_{ab} -\frac{\kappa_1^2}{16 b^2 \lambda^2}   \right) \Psi = 0 , \label{kswdw} 
\end{align}
which are Hermitian under the measure $\mu = 8 \sqrt{2} a^2 b^2$. We study each subalgebra separately.

\subsubsection{Subalgebra $(\hat{Q}_2, \hat{Q}_3)$}
In this case, eqs \eqref{kswdw}, \eqref{ksq2} and \eqref{ksq3} are solved and give the wave function
\begin{align}
\Psi (a,b)&= A \exp \left( i \ ab \cosh \frac{\lambda \ln a}{\kappa_1} \left(c_2 - \sqrt{16 + c_2^2} \tanh \frac{\lambda \ln a}{\kappa_1} \right)\right). 
\end{align}
We now perform the semiclassical analysis in order to draw some conclusions on the nature of the singularity. The quantum potential in this case is zero and therefore, as discussed in the introduction, the semiclassical solution is expected to be the same as the classical one. Indeed after solving the semiclassical system 
\begin{align}
\frac{b}{\kappa_1} \left( (-\lambda \sqrt{16 + c_2^2} +c_2 \kappa_1) \cosh \frac{\lambda \ln a}{\kappa_1} + (\lambda c_2 -\sqrt{16 +  c_2^2} \kappa_1) \sinh \frac{\lambda \ln a}{\kappa_1}
\right) &= \frac{32 \kappa_3^2 b^2 \dot{a}}{\kappa_1^2 n} + \frac{8ab \dot{b}}{n},\\
a \cosh \frac{\lambda \ln a}{\kappa_1} \left( c_2 -\sqrt{16+ c_2^2} \tanh \frac{\lambda \ln a}{\kappa_1} \right) &=\frac{8 ab \dot{a}}{n} + \frac{8a^2 \dot{b}}{n}, 
\end{align}
under the gauge choice $a=e^t$ we find that this is the case here. We proceed to the next subalgebra.
\subsubsection{Subalgebra $\hat{Q}_1$}
For the one-dimensional subalgebra $\hat{Q}_1$, the solution of eqs \eqref{ksq1} and \eqref{kswdw} is
\begin{align}
\Psi & = e^{-i c_1 \ln a} \left( A_1 J_{\frac{i c_1 \kappa_1}{\lambda}} (-i 4ab) + B_1 Y_{\frac{i c_1 \kappa_1}{\lambda}} (-i 4ab) \right),
\end{align}
where $J_\nu (z), Y_\nu (z)$ are the Bessel function of the first and second kind respectively. To bring this wave function in polar form, we consider approximation limits for the Bessel functions. The wave function for small and large arguments takes the form
\begin{align}
\Psi_{sm} &\approx D_1 \cos \left(\frac{c_1 \kappa_1}{\lambda} \ln (4ab)\right) e^{- i c_1 \ln a}, \\
\Psi_{la} & \approx D_2 \frac{1}{\sqrt{ab}} \sinh (4 ab + \frac{c_1 \kappa_1 \pi}{2 \lambda}) e^{-i c_1 \ln a}.
\end{align}
The quantum potential does not vanish and equals to $\mathcal{Q}_{sm} = -\frac{c_1^2 \kappa_1^2}{16 a^2 b^2 (\kappa_1^2 - 4 \kappa_3^2)}$ for small arguments of the Bessel function and $\mathcal{Q}_{la} = 1+ \frac{1}{64 a^2 b^2}$ for late times, therefore it is expected that the semiclassical solution will differ from the classical. Indeed, the semiclassical equations are explicitly written as
\begin{align}
\frac{32 \kappa_3^2 b^2 \dot{a}}{\kappa_1^2 n} + \frac{8 a b \dot{b}}{n} &=- \frac{c_1}{a} ,\\
\frac{8 a b \dot{a}}{n} + \frac{8 a^2 \dot{b}}{n} &=0
\end{align}
and by selecting the gauge $n=1$, the solution becomes 
\begin{align}
a  = d_2 \exp \left( \frac{c_1 \kappa_1^2 t}{8 d_1^2 (\kappa_1^2 - 4 \kappa_3^2)} \right), \quad
b= \frac{d_1}{c_1} \exp \left( \frac{c_1 \kappa_1^2 t}{8 d_1^2 (\kappa_1^2 - 4 \kappa_3^2)} \right)
\end{align}
where $d_i$ are integration constants. After proper re--parameterizations and change of coordinates we find that the line element is
\begin{align}
ds^2& =-\frac{\alpha}{T}dT^2 + \frac{1}{T} dr^2 + T d\theta^2 + T \sin^2 \theta d\varphi^2,
\end{align}
where $\alpha=\frac{4 d_1^2 \lambda^4}{c_1^2 \kappa_1^4}$. This spacetime has a singularity at $T \rightarrow 0$, which however disappears for $\alpha = \frac{1}{4}$ since the Ricci scalar becomes constant at this value
\begin{equation}\label{ricci_kantowski}
R = \frac{1-4 \alpha}{2 \alpha T}
\end{equation}

\section{Massless field in the FLRW universe}\label{Massless field in the closed FLRW universe}
\subsection{Classical treatment}
We now turn to the study of FLRW geometries. These belong to the Bianchi classification; note, however, that they appear as special solutions of several Bianchi types. The spatially closed case belongs to Bianchi type IX model, the spatially open to type V while the spatially flat to the Bianchi type I. We solve in detail the $k \neq 0$ cases and only mention the final result for the spatially flat case. The general form of the line element of a FLRW spacetime is
\begin{equation}\label{FLRW_metric}
ds^2 = - N^2(t) dt^2 +a^2 (t)\left(\frac{dr^2}{1-  kr^2} +r^2  d \theta^2 +r^2 \sin^2 \theta d\varphi^2\right), 
\end{equation}
where $N(t)$ is the lapse function and $a(t)$ is the scale factor. For this spacetime, the total Lagrangian for gravity plus matter system is
\begin{equation}
L=  6Nk  a -\frac{6a\dot{a}^2}{N} + \frac{a^3 \dot{\phi}^2}{2N},
\end{equation}
where a term of a total time derivative has been discarded. This Lagrangian has the singular form of equation \eqref{singular_lagrangian} and the procedure of integration of the system via the conditional symmetries applies here. To this end, we have to turn to the constant potential parametrization\footnote{For $k=0$, the Lagrangian is already in the constant potential parametrization. This results in the existence of numerous rheonomic integrals of motion corresponding to the infinite number of conformal Killing fields in two dimensions.} by setting $ n = 6 k N a$ resulting in
\begin{equation}\label{FLRWconstantpot}
L = n - \frac{36k  a^2 \dot{a}^2}{n} + \frac{3k  a^4 \dot{\phi}^2}{n},
\end{equation}
with the corresponding Hamiltonian constraint and supermetric respectively
\begin{equation}
\mathcal{H}= - \frac{p_a^2}{72 k  a^2} + \frac{p_\phi^2}{12 k   a^4} -1\approx 0, \quad
G_{\alpha \beta}= 6k  a  \label{FLRWsupermetric}
\left(\begin{array}{c c}
-12a & 0 \\
0 & a^3
\end{array}\right).
\end{equation}
This supermetric represents a flat two-dimensional space, admitting the following three symmetries and the homothecy
\begin{align}
\xi_1 = \frac{e^{\phi/\sqrt{3}}}{a}\partial_a-\frac{2 \sqrt{3} e^{\phi/\sqrt{3}}}{a^2}\partial_\phi, \quad
\xi_2 = \frac{e^{-\phi/\sqrt{3}}}{a}\partial_a + \frac{2 \sqrt{3} e^{-\phi/\sqrt{3}}}{a^2} \partial_\phi, \quad
\xi_3 = \partial_\phi, \quad
\xi_h = \frac{a}{4}\partial_a
\end{align}
where as before the numbered indices denote the Killing fields while $h$ denotes the homothetic field. These symmetry generators satisfy the following Lie bracket algebra
\begin{align}\label{poisson_massless_nonflat_FLRW}
[\xi_1,\xi_3]= -\frac{1}{\sqrt{3}}\xi_1, \quad
[\xi_2,\xi_3]= \frac{1}{\sqrt{3}} \xi_2, \quad
[\xi_1,\xi_h]= \frac{1}{2} \xi_1, \quad
[\xi_2,\xi_h]= \frac{1}{2}\xi_2,
\end{align}
For the case $k=0$ the corresponding algebra is the same but with structure constant coefficients being $C^1_{31} = C^2_{23} =\frac{\sqrt{3}}{4}, \ C^1_{1h} =C^2_{2h} =\frac{1}{2}$.
The corresponding first integrals of motion in the configuration space are:
\begin{align}
Q_1 &= - \frac{12 e^{\frac{\phi}{\sqrt{3}}} k   a \left( 6 \dot{a} + \sqrt{3} a \dot{\phi}\right) }{n}, \
Q_2   =  \frac{12 e^{-\frac{\phi}{\sqrt{3}}} k   a \left(- 6 \dot{a} + \sqrt{3} a \dot{\phi}\right) }{n} ,\\
Q_3 &=  \frac{6 k  a^4 \dot{\phi}}{n} ,\
Q_h = -\frac{18 k  a^3 \dot{a}}{n} + \int dt \ n(t) ,
\end{align}
In order to determine the line element, the system $Q_i=\kappa_i$ for $i=1,2,3,h$ with $\kappa_i$ constants is solved. The solution is 
\begin{align}
a&=  \frac{2 \times 3^{1/4} \sqrt{\kappa_3} e^{\frac{\phi}{2 \sqrt{3}}}}{\sqrt{-\kappa_1 + \kappa_2 e^{\frac{2 \phi}{\sqrt{3}}}}} ,\\
\dot{a} &=-\frac{\sqrt{\kappa_3} e^{\frac{\phi}{2 \sqrt{3}}} (\kappa_1 + \kappa_2 e^{\frac{2 \phi}{\sqrt{3}}}) \dot{\phi}}{3^{1/4} (-\kappa_1 + \kappa_2 e^{\frac{2 \phi}{\sqrt{3}}})^{3/2}},\\
n &=  \frac{288 \kappa_3 e^{\frac{2 \phi}{\sqrt{3}}} k  \dot{\phi}}{(\kappa_1 - \kappa_2 e^{\frac{2 \phi}{\sqrt{3}}})^2} ,\\
\int dt \ n(t)  &= -\kappa_h -\frac{\sqrt{3} \kappa_3 (\kappa_1 + \kappa_2 e^{\frac{2 \phi}{\sqrt{3}}})}{2 (\kappa_1 -\kappa_2 e^{\frac{2 \phi}{\sqrt{3}}})},
\end{align}
while a relation for the constants appearing in the system is also found by the only non trivial consistency condition $\frac{\partial}{\partial t}\int dt \ n(t) =n $,
\begin{equation}\label{massless_flrw_constants}
\kappa_1 \kappa_2 + 144k  = 0.
\end{equation}
This relation is the constraint equation of the theory, and it is interesting to note that the Casimir invariant of the algebra is the term $Q_1 Q_2$ which is also the kinetic part of the Hamiltonian.

The system of equations is $Q_i = \kappa_i$, where $i = 1,2,3,h$ is solved by setting one of the constants $\kappa_1, \ \kappa_2$ equal to zero. This is necessary because of the vanishing of the constraint and the fact that its kinetic part ($k=0$) when  expressed with respect to $Q_i$'s equals $Q_1 Q_2$. As before, it is also here true that this product is the Casimir invariant of the algebra.

In this form, the solution still contains an arbitrary function of time, representing the time reparametrization invariance (since no gauge fixing has been assumed in deriving the solution). Choosing the gauge $\phi = \ln t$ we find that the solution is
\begin{align}
a = \frac{2 \times 3^{1/4} \sqrt{\kappa_3} t^{\frac{1}{2 \sqrt{3}}}}{\sqrt{-\kappa_1 + \kappa_2 t^{2/\sqrt{3}}}}, \quad
N = \frac{8 \times 3^{3/4} \sqrt{\kappa_3} t^{-1 + \frac{\sqrt{3}}{2}}}{(-\kappa_1 + \kappa_2 t^{2/\sqrt{3}})}.
\end{align}
By inserting it in the line element \eqref{FLRW_metric} and performing proper coordinate transformations in order to absorb the redundant constants, the final line element of the spacetime is
\begin{align}
ds^2  = -\frac{\lambda}{4 \sqrt{T} (1+T \epsilon)^3} dT^2 + \frac{\lambda \sqrt{T}}{(1+T \epsilon)} \left( \frac{dr^2}{1- r^2 \epsilon} + r^2 d\theta^2 + r^2 \sin^2 \theta d\varphi^2 \right).
\end{align}
This geometry has one essential constant, as can be shown by using the methodology of \cite{Papadopoulos:2005ft,Christodoulakis:2003wy}. The Ricci scalar is 
\begin{equation}
R =-\frac{3 (T\epsilon + 1)^3}{2 T^{3/2} \lambda}, 
\end{equation}
where we have set $\lambda = - \frac{\kappa_3}{\sqrt{3} k^{3/2}}$ rendering the metric element singular for both $T \rightarrow 0$ and $T \rightarrow \infty$. 
The same steps for the spatially flat case lead us to the solution
\begin{align}
ds^2  =-dT^2 + T^{2/3} dr^2 + T^{2/3}r^2 d\theta^2 + T^{2/3}r^2 \sin^2 \theta d\varphi^2.
\end{align}
This spacetime metric is conformally flat with Ricci scalar 
\begin{equation}
R = -\frac{2}{3 T^2}.
\end{equation}
It does not contain any essential constants characterizing the geometry and the matter content of the spacetime and, as can be seen from the form of the Ricci scalar, a singularity for $T \rightarrow 0$ appears.
\subsection{Canonical quantization and semiclassical analysis}\label{quantum}
We next canonically quantise the classical system. Promoting the constraints as well as the first integrals $Q_i$ to operators and imposing them on the wave function gives the following quantum equations:
\begin{align}
\hat{Q}_1 \Psi &= -\frac{i e^{\phi /\sqrt{3}} (-6 \partial_\phi \Psi + \sqrt{3} a \partial_a \Psi)}{\sqrt{3} a^2} = \kappa_1 \Psi, \label{q1flrwclosed} \\
\hat{Q}_2 \Psi &= -\frac{i e^{-\phi /\sqrt{3}} (6 \partial_\phi \Psi + \sqrt{3} a \partial_a \Psi)}{\sqrt{3} a^2} = \kappa_2 \Psi, \label{q2flrwclosed} \\
\hat{Q}_3 \Psi &= - i \partial_\phi \Psi = \kappa_3 \Psi, \label{q3flrwclosed}\\
\hat{\mathcal{H}} \Psi &= \frac{-144 k  a^4 \Psi - 12 \partial_{\phi\phi} \Psi + a (\partial_a \Psi + a \partial_{aa} \Psi)}{ 144 k  a^4}=0, \label{wdwflrwclosed}
\end{align}
where the measure is $\mu (a,\phi) = 6 \sqrt{3} a^3 k $. The quantum equations $\hat{Q}_i \Psi= \kappa_i \Psi$ that can be imposed simultaneously according to the condition \eqref{selection_rule} are the two-dimensional $(\hat{Q}_1, \hat{Q}_2)$ and the one-dimensionals $\hat{Q}_1,\hat{Q}_2,\hat{Q}_3$. The one-dimensional subalgebras spanned by the operators $\hat{Q}_1,\hat{Q}_2$ give solutions which are special cases of the two-dimensional case \cite{Christodoulakis:2012eg}.

\subsubsection{Subalgebra $(\hat{Q}_1 , \hat{Q}_2)$}
For the case of the two-dimensional subalgebra, we solve the eqs \eqref{q1flrwclosed}, \eqref{q2flrwclosed} and \eqref{wdwflrwclosed}. The solution for the wave function is
\begin{equation}
\Psi= A \exp \left(i \frac{a^2}{4} (\kappa_1 e^{-\frac{\phi}{\sqrt{3}}} + \kappa_2 e^{\frac{\phi}{\sqrt{3}}})\right).
\end{equation}
The semiclassical analysis is next performed following Bohm as explained in the introduction. This wave function is written in polar form and we can see that the amplitude $\Omega$ is constant. Therefore, the quantum potential will vanish rendering the solution for this case same as the classical metric. Indeed, if we solve the semiclassical solutions 
\begin{align}
\frac{1}{2} a \left( -e^{-\phi/\sqrt{3}} (\kappa_1 + e^{2\phi/\sqrt{3}} \kappa_2)  \right) & = \frac{144 k  a \dot{a}}{n},\\
\frac{a^2}{12} \left( \sqrt{3} e^{-\phi/\sqrt{3}} (\kappa_1 -e^{2\phi/\sqrt{3}} \kappa_2) \right) &= -\frac{72 k  a^2 \dot{\phi}}{n},
\end{align}
with phase function $S =\frac{1}{4} a^2 e^{-\frac{\phi}{\sqrt{3}}} (\kappa_1 +\kappa_2 e^{\frac{2 \phi }{\sqrt{3}}})$ we indeed find the same line element as in the classical case. The same conclusion also holds for $k=0$.

\subsubsection{Subalgebra $\hat{Q}_3$}
In the case of the one-dimensional algebra, the system of equations is formed by the equations \eqref{wdwflrwclosed} and \eqref{q3flrwclosed}. The wave function is
\begin{align}
\Psi_{cl} (a,\phi) &= e^{i \phi \kappa_3} (A_1 I_{-i \sqrt{3} \kappa_3} (6 a^2 ) + B_1  I_{i \sqrt{3} \kappa_3} (6 a^2 )),  \\
\Psi_{op} (a,\phi) &= e^{i \phi \kappa_3} (A_2 J_{-i \sqrt{3} \kappa_3} (6 a^2 ) + B_2  J_{i \sqrt{3} \kappa_3} (6 a^2 )),
\end{align}
for the closed and open case respectively. In order to write the wave function in polar form for the semiclassical analysis, approximation limits are taken, for small and large arguments of the Bessel functions. Using the simplifying assumption $A_1 = B_1, \ A_2=B_2$ renders the, common for the two cases, wave function
\begin{equation}
\Psi_{sm} \approx c_1 e^{i \kappa_3 \phi} \cos \ln a.
\end{equation}
Similarly for the large values, assuming again $A_1 = B_1, \ A_2=B_2$ and using the formulas of the appendix the wave function becomes
\begin{equation}
\Psi^{cl}_{la} \approx \frac{e^{a^2}}{a} e^{i \kappa_3 \phi}, \quad 
\Psi^{op}_{la} \approx \frac{\sin (6 a^2)}{a} e^{i \kappa_3 \phi}.
\end{equation}
The quantum potential for small values does not vanish $\mathcal{Q}_{sm}= \frac{1}{144 k  a^4} $, while $\mathcal{Q}_{la}^{cl} = - \frac{1+ 4 a^4}{144 a^4 k }$ and $\mathcal{Q}_{la}^{op} = \frac{144 k a^4 -1}{144 k  a^4}$. The phase function is $S= \kappa_3 \phi$. The solution of the semiclassical equations with respect to $(a,n)$ is 
\begin{align}
a = c, \quad n = \frac{6 k a^4}{\kappa_3} \dot{\phi},
\end{align}
and has a remaining freedom for the scalar field which we select to be such that the lapse function $N(t)$ of the semiclassical element is the same as for the classical, that is
\begin{align}
\phi (t) = - \frac{8 \times 3^{3/4} t^{{\sqrt{3}/2}}  \sqrt{- \frac{48 k  t^{2/\sqrt{3}}}{\kappa_1}-\frac{\kappa_1}{3}} \kappa_1 \kappa_3^{3/2}  \left( -3 + \sqrt{1+ \frac{144k  t^{2/\sqrt{3}} }{\kappa_1^2}}
\, _2F_1\left(\frac{1}{2},\frac{3}{4};\frac{7}{4};-\frac{144 k  t^{\frac{2}{\sqrt{3}}}}{\kappa_1^2}\right)
 \right)}{c^3 (144 k  t^{2/\sqrt{3}} \kappa_1 + \kappa_1^3)},
\end{align}
where $\, _2F_1\left(a,b;c;d \right)$ is the Gauss hypergeometric function. Inserting the solution in the 4-dimensional element and after proper coordinate transformations the spacetime metric is written
\begin{equation}
ds^2= -\frac{\lambda}{4\sqrt{T} (1+T \epsilon)^3} dt^2 + \frac{1}{1- \epsilon r^2} dr^2 + r^2 d\theta^2 +r^2 \sin^2 \theta d\varphi^2,
\end{equation}
where the sign $(+)$ accounts for the closed case and $(-)$ for the open case while the identification $c^2= \frac{ \lambda^2}{16}$ has been considered in order for the constant $\lambda$ coincide with that one in the classical metric. This spacetime has the interesting property of having constant Ricci scalar $R= 6 k $, all higher derivatives of its Riemann tensor zero and constant all curvature scalars constructed from its Riemann tensor. Hence, there is no curvature and/or higher derivative curvature singularity. 

Following the same procedure for the spatially flat case, we find that the wave function is of the form
\begin{align}
\nonumber \Psi (a,\phi) &= e^{i \kappa_3 \phi} \left(A_3 \cos (2 \sqrt{3} \kappa_3 \ln a) + B_3 \sin (2 \sqrt{3} \kappa_3 \ln a) \right).
\end{align}
Note that in this case there is no need to make an approximation in order to write it in polar form as it is already in this form with $\Omega = A_3 \cos (2 \sqrt{3} \kappa_3 \ln a) + B_3 \sin (2 \sqrt{3} \kappa_3 \ln a)$ and $S = \kappa_3 \phi$. The line element we obtain is
\begin{equation}
ds^2 = - dT^2 + dr^2 + r^2 d\theta^2 + r^2 \sin^2 \theta d\varphi^2.
\end{equation}
This spacetime is the Minkowski spacetime and does not contain any essential constants. Therefore, there is no singularity for the range of all times, since no approximation limits have been considered here.


\section{Bianchi type II coupled to a massless scalar field}\label{bianchi_II}
\subsection{Classical analysis}
The Bianchi type II spacetime metric has the general form of eq. \eqref{bianchi_metric} with invariant one-forms being
\begin{equation}
\sigma =\left(
\begin{array}{ccc}
 0 & 1 & -x \\
 0 & 0 & 1 \\
 1 & 0 & 0 \\
\end{array}
\right).
\end{equation}
We choose the $\gamma$ matrix to have the diagonal form $\gamma= diag (e^{2a}, e^{2b} ,e^{2c})$ where $a,b,c$ are scale factors. Therefore, the 4-dimensional line element becomes
\begin{equation}
ds^2 = -N^2 dt^2 + e^{2c} dx^2 + e^{2a} dy^2 -2 e^{2a} x dy dz +( e^{2a} x^2 + e^{2b}) dz^2.
\end{equation}
The integrand of the total action \eqref{total_action}, after discarding a term of total derivative, becomes 
\begin{equation}
L= -\frac{1}{2N}e^{a+b+c}\left(4\dot{b}\dot{c}+4\dot{a}\dot{b}+4\dot{a}\dot{c}-\dot{\phi}^2 \right)-\frac{N}{2}e^{3a-b-c}.
\end{equation}
We choose to apply the conditional symmetries method in the constant potential parametrization by setting the lapse equal to $N=2n\exp\left(-3a+b+c\right)$ where $n$ denotes the lapse function in the new parametrization. The new Lagrangian is therefore
\begin{equation}\label{transformed_lag_bianchi_II}
L=-\frac{1}{4n}e^{4a}\left(4\dot{a}\dot{b}+4\dot{a}\dot{c}+4\dot{b}\dot{c}-\dot{\phi}^2 \right)-n,
\end{equation}
from which we can read the supermetric $G_{\alpha \beta}$ in these 
\bal
G_{\alpha \beta} =e^{4a}
\begin{pmatrix}
 0 & -1 & -1 & 0 \\
-1 &  0 & -1 & 0 \\
-1 & -1 &  0 & 0 \\
 0 &  0 &  0  & \frac{1}{2}
\end{pmatrix}.
\eal
This superspace is conformally flat and admits six Killing vector fields $\xi_i,\, i=1,\dots,6$ and one homothetic vector field $\xi_h$, which are
\begin{align}
\xi_1 &=(a+b)\p_b-(a+c)\p_c, \quad \xi_2 =\frac{1}{2}\phi\p_c+(a+b)\p_\phi,  \quad \xi_3 = \p_b, \non\\
\xi_4 &= \phi \p_b +2(a+c) \p_\phi,  \quad \xi_5 =\p_c \quad \xi_6  =\p_\phi \quad \xi_h  =\frac{1}{4}\p_a.
\end{align}
The non vanishing commutators of the Lie algebra spanned by the above vector fields are
\begin{align}\label{poisson_bianchi_ii}
[\xi_1,\xi_2] &=\xi_2,    \quad  [\xi_1,\xi_3] =-\xi_3,   \quad [\xi_1,\xi_4] =-\xi_4,  \quad [\xi_1,\xi_5] =-\xi_5, \non\\
[\xi_2,\xi_3] &= -\xi_6,  \quad  [\xi_2,\xi_4] = \xi_1,  \quad  [\xi_2,\xi_6] = -\frac{1}{2}\xi_5,  \quad  [\xi_4,\xi_5] = -2\xi_6 \non\\
[\xi_4,\xi_6] &= -\xi_3,  \quad  [\xi_1,\xi_h] = -\frac{1}{2}(\xi_3-\xi_5)  \quad  [\xi_2,\xi_h] =-\frac{1}{2} \xi_6, \quad [\xi_4,\xi_6] =-\xi_6.
\end{align}
The six first integrals of motion $Q_i$, in the velocity phase space, generated by the corresponding $\xi_i, i=1,\dots, 6$ are 
\bal
Q_1 &=\frac{e^{4a}}{n}\left((\dot{a}+\dot{b}) c+(\dot{b}-\dot{c}) a - (\dot{a}+\dot{c}) b \right), \quad 
Q_2 =\frac{e^{4a}}{2n}\left(-(\dot{a}+\dot{b}) \phi+(a+b) \dot{\phi} \right), \quad
Q_3 =-\frac{e^{4a}}{n}\left(\dot{a}+\dot{c}\right), \non \\
\quad Q_4 &=\frac{e^{4a}}{n}\left(-(\dot{a}+\dot{c}) \phi+(a+c) \dot{\phi} \right), \quad
Q_5 =-\frac{e^{4a}}{n}\left(\dot{a}+\dot{b}\right), \quad 
Q_6 =\frac{e^{4a}}{2n}\dot{\phi},
\eal
while the rheonomic integral of motion arising from the vector field $\xi_h$ is
\bal
Q_h=-\frac{e^{4a}}{4n}\left(\dot{b}+\dot{c} \right)-\int\! dt\, n.
\eal
The solution of the system $Q_i = \kappa_i, i=1,\dots, 6$ is
\bal
n =-\frac{e^{4a}}{\kappa_5}\left(\dot{a}+\dot{b} \right), \quad
\phi & =-\frac{2\kappa_6}{\kappa_5}(a+b)+\frac{2\kappa_2}{\kappa_5}, \quad
c  =\frac{\kappa_3}{\kappa_5}\left( a+b \right)-a+\frac{\kappa_4 \kappa_5-2\kappa_2 \kappa_3}{2\kappa_5 \kappa_6},
\eal
along with the consistency equation
\bal
2\kappa_1 \kappa_6+\kappa_4 \kappa_5-2\kappa_2 \kappa_3=0.
\eal
In order to determinate the final function $b$ we insert the above functions into the integral $Q_h$ and we end up with
\bal
b=c_2-a-\frac{1}{2}c_1\epsilon \arctan\left(\frac{\sqrt{\kappa_5-c_1^2 e^{4a}}}{\kappa_5}\right),
\eal
where $\epsilon=\pm 1$ and $c_1,c_2$ are constants of integration. The various constants that appear in the solution are constrained by the nihilism of the Hamiltonian constraint $\mathcal{H}$, resulting to $\kappa_5^2+c_1^2 (\kappa_6^2-\kappa_3 \kappa_5)=0$. 

The spacetime metric, after making the gauge choice $a=\frac{1}{4}\ln\left( (\kappa_3 \kappa_5-\kappa_6^2)\, \sech^2 t\right)$ and removing the inessential constants, is
\bal \label{BianchiIIcl}
ds^2 =& -e^{(\alpha+\nu) t} \cosh t \, dt^2 + e^{-\alpha t}\cosh t \, dx^2  +\sech t \,dy^2 -2x\, \sech t\, dy\, dz + \left(e^{-\nu t} \cosh t + x^2\,\sech t\right) dz^2
\eal
In this line element the singularity appears for $t= \pm \infty$ (depending on the values of the constants $\alpha,\nu$) as can be seen from the Ricci scalar
\bal
R=-\frac{1}{2}e^{(\alpha+\nu)t}(\alpha \nu-1)\,\sech t.
\eal
\subsection{Quantization and semiclassical analysis}
For this model, the admissible subalgebras compatible with the condition \eqref{selection_rule} are the three-dimensional $(\hat{Q}_3, \hat{Q}_5, \hat{Q}_6)$, the two-dimensionals $(\hat{Q}_1, \hat{Q}_6)$, $(\hat{Q}_3, \hat{Q}_4)$, $(\hat{Q}_2, \hat{Q}_5)$ as well as all the one-dimensional ones for which we omit the detailed analysis since they all belong to the higher-dimensional algebras. The system of the quantum equations are
\begin{align}
\hat{Q}_1 \Psi &= i ((a+c) \partial_c - (a+b)\partial_b) \Psi, \\
\hat{Q}_2 \Psi &= -\frac{i}{2} (2 (a+b) \partial_\phi + \phi \partial_c) \Psi, \label{q2bianchi_II}\\
\hat{Q}_3 \Psi & = - i \partial_b \Psi, \\
\hat{Q}_4 \Psi & = - i (2 (a+c) \partial_\phi + \phi \partial_b) \Psi, \label{q4bianchi_II}\\
\hat{Q}_5 \Psi & = - i \partial_c \Psi, \label{q5bianchi_II}\\
\hat{Q}_6 \Psi & = - i \partial_\phi \Psi, \\
\hat{\mathcal{H}} \Psi & = -\frac{1}{4} e^{-4a} ((4 e^{4a}-8) + 4 \partial_{\phi \phi} - 4 \partial_c + \partial_{cc} -4 \partial_b - 2 \partial_{bc} + \partial_{bb} + 4 \partial_a -2 \partial_{ac} -2 \partial_{ab} + \partial_{aa}) \Psi, \label{wdwbianchi_II}
\end{align}
where the measure under which the operators are hermitian is taken to be equal to $\mu = e^{8a}$ (and of course assuming suitable boundary conditions). We proceed to the quantization for each subalgebra and the semiclassical analysis.
\subsubsection{Subalgebra $(\hat{Q}_3, \hat{Q}_5, \hat{Q}_6)$}
The solution for this case is 
\begin{align}
\Psi (a,b,c,\phi) =
\left( A_1 J_{-\lambda} (e^{2a} )+ A_2 J_\lambda (e^{2a} ) \right) e^{i S},
\end{align}
where 
\begin{equation}
S = \kappa_3 b + \kappa_5 c + (\kappa_3 + \kappa_5) a + \kappa_6 \phi,
\end{equation}
is the phase function and $\lambda^2 =3-\kappa_3 \kappa_5 + \kappa_6^2$. In order to perform the semiclassical analysis, the wave function has to be written in polar form. In order to do this, we study the behaviour of the argument of the Bessel functions. Depending on the values of the constants, it can be selected to become very large thus corresponding to the large arguments of the Bessel function. For the case of the small arguments, the approximation, assuming that $A_1 = A_2$, gives us for the wave function
\begin{equation}
\Psi_{sm} \approx e^{iS} e^{-2a} \cosh (2 \lambda a).
\end{equation}
The quantum potential in this case is $\mathcal{Q}_{sm} = e^{-4a} (2-\kappa_3 \kappa_5 + \kappa_6^2)$. 
For the large arguments, the same approximation gives the wave function
\begin{equation}
\Psi_{la} \approx e^{iS} e^{-3a} (\cos e^{2a} +  \sin e^{2a}) 
\end{equation}
and we have for the quantum potential 
\begin{equation}
\mathcal{Q}_{la} = -1 - \frac{3 e^{-4a}}{4}
\end{equation}
Solving the semiclassical equations $\frac{\partial L}{\partial \dot{q_i}} = \frac{\partial S}{\partial q_i}$ for $a,b,c,n$, making a convenient gauge choice for the scalar field and absorbing the inessential constants by making allowed coordinate transformations, the line element is finally takes the form
\begin{align}
ds^2 = - \lambda_1 e^T dT^2 + T^2 dx^2 + dy^2 -2x dy dz + \frac{1}{4} \left(4 x^2 + (\lambda_2 T + \lambda_3 )^2 \right) dz^2
\end{align}
where $\lambda_1, \ \lambda_2, \ \lambda_3 $ are combinations of constants. The Ricci scalar is 
\begin{equation}
R = -\frac{2}{T^2 (\lambda_2 T + \lambda_3)} - \frac{2 \lambda_2^2 T -2 \lambda_2^2 + 3 \lambda_2 \lambda_3}{\lambda_1 e^T (\lambda_2 T + \lambda_3)} + \frac{2 \lambda_2 \lambda_3 - \lambda_3^2}{\lambda_1 e^T T (\lambda_2 T + \lambda_3)}
\end{equation}
This metric has a singularity at $T \rightarrow 0$.

\subsubsection{Subalgebra $(\hat{Q}_1,\hat{Q}_6)$}
For this subalgebra, as well as for all other two-dimensional subalgebras, the Wheeler-DeWitt and the eigenvalue equations for the members of the subalgebra in question are not enough to give a unique solution for the wave function. This being the case, we can turn in the Casimir invariant of the algebra and promote it to operator, an action which is permissible in view of the fact that it commutes with all the members of the full algebra. The Casimir invariant is 
\begin{align} \label{casbianchi_II}
\hat{Q}_{cas} = 4 \hat{Q}_1 \hat{Q}_6 -2 \hat{Q}_2 \hat{Q}_3 -2 \hat{Q}_3 \hat{Q}_2 + \hat{Q}_4 \hat{Q}_5 +\hat{Q}_5 \hat{Q}_4 + A (\hat{Q}_3 \hat{Q}_5 - \hat{Q}_6^2),
\end{align}
where $A$ is a constant. This gives the eigenvalue equation with constant value $\kappa_{cas}$ that is imposed on the wave function. Thus we have the additional equation
\begin{align}
\hat{Q}_{cas} \Psi = A (\partial_{\phi \phi} -\partial_{bc}) \Psi = \kappa_{cas} \Psi.
\end{align}
Now, the solution of the system of equations \eqref{q2bianchi_II}, \eqref{q4bianchi_II}, \eqref{wdwbianchi_II} and \eqref{casbianchi_II} is
\begin{align}\label{16wavefunction}
\Psi = e^{i S} \left( 
I_{ i \kappa_1} (2 \sqrt{-(a+b) (a+c) \kappa_6^2}) + K_{i \kappa_1} (2 \sqrt{-(a+b) (a+c) \kappa_6^2}) \right) \left( A_1 J_{-\sqrt{3}} (e^{2a}) + A_2 J_{\sqrt{3}} (e^{2a}) \right)
\end{align} 
where 
\begin{align}
S = \kappa_1 \ln (a+b) + \kappa_6 \phi - \kappa_6^2 \frac{\kappa_1}{2} \ln \left(-(a+b) (a+c) \right)
\end{align}
is the phase function which we identify with the action. In order to write the wave function in polar form we make approximations so that the Bessel arguments take small values, thus
\begin{equation}
\Omega_{sm} =  e^{-2a} \cos (\kappa_1 \ln (2 \sqrt{- (a+b) (a+c) \kappa_6^2})) \cosh (2 \sqrt{3} a)
\end{equation}
which gives a quantum potential equal to $\mathcal{Q}_{sm}  \approx \frac{(8 a^2 + 8 bc + 8a b + 8ac + \kappa_1^2)}{4 e^{4a} (a+b) (a+c)}$. For the large arguments, the phase function remains the same as in the last case while the coefficient $\Omega$ becomes 
\begin{equation}
\Omega_{la} = e^{-3a +2 \sqrt{-(a+b) (a+c) \kappa_6^2}} \frac{\cos e^{2a} + \sin e^{2a}}{\sqrt{-(a+b) (a+c) \kappa_6^2}}
\end{equation}
which renders the quantum potential 
$\mathcal{Q}_{la} \approx \frac{\left( -1 + 2 \sqrt{-(a+b) (a+c) \kappa_6^2} - (a+b) (a+c) (3+ 4 e^{4a} -4 \kappa_6^2)\right)}{4 e^{4a} (a+b) (a+c)} $. Since the quantum potentials do not vanish, it is expected that the semiclassical spacetime will differ from the classical. Following the usual steps, the final line element is 
\begin{align}
ds^2 = -\frac{\lambda_1^2}{(1+T)^2} e^{2 c_1 T + \frac{2 (c_1^2 +c_2)}{c_1 (1+T)}} dT^2 + T^2 dx^2 + dy^2 -2 x dydz + \left( \frac{(c_2 -c_1^2 T)^2}{c_1 + c_1 T} + x^2 \right) dz^2 
\end{align}
in which remain three essential constants after performing allowed coordinate transformations. From the Ricci scalar we find that the singularity appears for $T \rightarrow 0$ and $T \rightarrow \infty $ when $c_1 <0$ while for $c_1 >0$ the Ricci scalar is zero, thus the singularity is resolved for this range of the $c_1$ parameter.
\subsubsection{Subalgebra $(\hat{Q}_2,\hat{Q}_5)$ and $({\hat{Q}_3, \hat{Q}_4})$}
The solution of the system for the subalgebra $(\hat{Q}_2,\hat{Q}_5)$ is given by the use of equations \eqref{q2bianchi_II}, \eqref{q5bianchi_II}, \eqref{wdwbianchi_II} and \eqref{casbianchi_II} is
\begin{align}\label{25wavefunction}
\Psi =  \left( c_1 J_{-\sqrt{3}} (e^{2a})  + c_2 J_{\sqrt{3}} (e^{2a})\right) e^{i S},
\end{align} 
where 
\begin{align}
S = \frac{4 \kappa_5^2 a^2 + 4 \kappa_5^2 bc +4 \kappa_5^2 (b+c) - (\kappa_5 \phi -2 \kappa_2)^2}{4 \kappa_5 (a+b)},
\end{align}
is the phase function. In order to write the wave function in polar form we make approximations so that the Bessel arguments take small values for the classical values of the scale factors. 
\begin{equation}
\Omega_{sm} =  e^{-2a} \cos (\sqrt{a+b}) \cosh(2 \sqrt{3}a),
\end{equation}
which gives a quantum potential equal to $\mathcal{Q}_{sm}  \approx 2 e^{-4a}$. For the large arguments, the phase function remains the same as in the last case while the coefficient $\Omega$ becomes 
\begin{equation}
\Omega_{la} = \frac{\cos e^{2a} + \sin e^{2a}}{e^{3a} \sqrt{a+b}},
\end{equation}
which renders the quantum potential 
$\mathcal{Q}_{la} \approx -1 - \frac{3}{4 e^{4a}} $. Since the quantum potentials do not vanish, it is expected that the semiclassical spacetime will differ from the classical. Following the usual steps for finding the semiclassical solution, we find that the final line element is 
\begin{align}
ds^2 = -\lambda_2^2 e^{\lambda_3 T} dt^2 + \lambda_1^2 T^2 dx^2 + dy^2 -2x dydz + \left(  x^2 + (1+T)^2\right) dz^2,
\end{align}
in which remain three essential constants after performing allowed coordinate transformations. The Ricci scalar becomes infinite for $T \rightarrow 0$ as well as for $T \rightarrow \infty$ for negative values of $\lambda_3$ parameter, while it vanishes at this limit for positive values of this parameter. 

The subalgebra $({\hat{Q}_3, \hat{Q}_4})$ is not studied separately because it has the same solution with this algebra under the replacements $\kappa_5 \rightarrow \kappa_3, \ \kappa_2 \rightarrow -\frac{\kappa_4}{2}$.


\section{Bianchi type V coupled to a massless scalar field}\label{bianchi_V}
\subsection{Classical analysis}
The appropriate invariant one-forms for this model are
\begin{equation}
\sigma =\left(
\begin{array}{ccc}
 0 & e^{-x} & 0 \\
 0 & 0 & e^{-x} \\
 1 & 0 & 0 \\
\end{array}
\right),
\end{equation}
while the $\gamma$ matrix assumes the diagonal form $\gamma = diag (a^4,b^4,c^2)$. The 4-dimensional spacetime is therefore described by
\begin{equation}
ds^2 = -N^2 dt^2 + c^2 dx^2 + e^{-2 x} a^4 dy^2 + e^{-2 x} b^4 dz^2 .
\end{equation}
This spacetime element, in order to give a valid Lagrangian, has to be simplified by taking into account the solution of the Einstein equation $G^1_2=T^1_2$. This equation gives the condition $c=ab$. The integrand of the total action \eqref{total_action}, after discarding a term of total derivative, becomes 
\begin{equation}
L = -6ab N -\frac{4ab^3 \dot{a}^2}{N} - \frac{16 a^2 b^2 \dot{a} \dot{b}}{N}-\frac{4a^3 b \dot{b}^2}{N} + \frac{a^3 b^3 \dot{\phi}^2}{2N} .
\end{equation}
In order to assume the constant potential parametrization, we make the transformation $N=\frac{n}{6ab}$, where $n$ is the reparametrized lapse function. The Lagrangian therefore becomes
\begin{equation}
L =-n -\frac{96 a^3 b^3 \dot{a} \dot{b}}{n}-\frac{24 a^2 b^4 \dot{a}^2}{n}-\frac{24 a^4 b^2 \dot{b}^2}{n}+\frac{3 a^4 b^4 \dot{\phi}^2}{n} .
\end{equation}
The symmetries of the system are more explicit in the new coordinates $(u,v,\phi)$ where 
\begin{equation}\label{coord_trans_bianchi_V}
a= e^{\frac{u+v}{4}}, \qquad b= e^{\frac{1}{4} (u-v)},
\end{equation}
while $\phi$ remains the same. The corresponding Hamiltonian constraint and the supermetric are
\begin{align}
\mathcal{H}& = 1 - \frac{1}{36} e^{-2u} p_u^2 + \frac{1}{12} e^{-2u} p_v^2 + \frac{1}{12} e^{-2u} p_\phi^2 \approx 0 ,
\end{align}
\begin{align}
G_{\alpha \beta} &= \left(
\begin{array}{ccc}
 -18 e^{2 u} & 0 & 0 \\
 0 & 6 e^{2 u} & 0 \\
 0 & 0 & 6 e^{2 u}  \\
\end{array}
\right). 
\end{align}
This superspace is conformally flat and has the following Killing fields and homothecy
\begin{align}
\xi_1 = \partial_v, \quad
\xi_2 =\phi \partial_v-v \partial_\phi,\quad
\xi_3 = \partial_\phi, \quad
\xi_h = \frac{1}{2} \partial_u,
\end{align}
satisfying the Lie algebra 
\begin{align}\label{poisson_bianchi_v}
[\xi_1,\xi_2]=- \xi_3, \quad
[\xi_2,\xi_3]= -\xi_1.
\end{align}
The first integrals of motion in the velocity phase space are
\begin{align}
Q_1 = \frac{6 e^{2 u} \dot{v}}{n}, \quad
Q_2 =  \frac{6 e^{2 u} \left(\phi  \dot{v}-v \dot{\phi}\right)}{n}, \quad
Q_3 = \frac{6 e^{2 u} \dot{\phi}}{n}, \quad
Q_h =-\frac{9 e^{2 u} \dot{u}}{n} - \int dt \ n(t) ,
\end{align}
and the solution of the system $Q_i = \kappa_i$ is
\begin{align}
n = \frac{6 e^{2u} \dot{v}}{\kappa_1}, \quad
u = \ln \left(  2 \sqrt{c_2} \sinh^{-1} \left( \frac{2 \sqrt{c_2} (c_3 \kappa_1 -2 v)}{\kappa_1} \right)  \right), \quad
\phi  = c_1 + \frac{\kappa_3 v}{\kappa_1},\quad
\kappa_2 = c_1 \kappa_1,
\end{align}
where the constants $c_i$ are constants of integration and the last relation is used to reduce the number of constants in our solution. A second relation for the constants appears due to the demand the solution satisfies the constraint equation, which is
\begin{equation}
\kappa_1^2+\kappa_3^2  =48 c_2.
\end{equation}
We make the choice of gauge 
\begin{equation}
v = \frac{1}{4} (-2 c_3 \kappa_1 - \frac{\kappa_1 \cosh^{-1} (\frac{1}{t})}{\sqrt{c_2}}) ,
\end{equation}
and return to the initial variables using the inverse transformation \eqref{coord_trans_bianchi_V}. After this procedure, the spacetime metric is found to be
\begin{equation}
ds^2 = -\frac{\sqrt{c_2}}{2 \sinh^3 T} dT^2 + \frac{2 \sqrt{c_2}}{\sinh T} dx^2 + \frac{2 e^{-2 x - \lambda T}}{\sinh T} dy^2 + \frac{2 e^{-2 x + \lambda T}}{ \sinh T} dz^2 ,
\end{equation}
where $\lambda=\frac{\kappa_1}{4 \sqrt{c_2}}$. This metric contains two essential constants. The singularity appears at $T \rightarrow \infty$ as can be seen from the Ricci scalar
\begin{equation}
R = \frac{(\lambda^2 - 3) \sinh^3 T}{\sqrt{c_2}} .
\end{equation}
and disappears for $\lambda^2 =3$.
\subsection{Quantization and semiclassical analysis}
For this model, the admissible subalgebras compatible with the condition \eqref{selection_rule} is the two-dimensional $(\hat{Q}_1,\hat{Q}_3)$ and the one-dimensional $\hat{Q}_2$. The quantum equations are
\begin{align}
\hat{Q}_1 \Psi &= - i \partial_v \Psi = \kappa_1 \Psi ,\\
\hat{Q}_2 \Psi &= i (v \partial_\phi - \phi \partial_v) \Psi = \kappa_2 \Psi ,\\
\hat{Q}_3 \Psi &= - i  \partial_\phi \Psi = \kappa_3 \Psi ,\\
\hat{\mathcal{H}} \Psi &= \frac{e^{-2u}}{144} \left(
(1+ 144 e^{2u})  -12 \partial_{\phi \phi} -12 \partial_{vv} + \partial_{u} +  \partial_{uu} \right) \Psi=0,
\end{align}
where the measure is $\mu = 9 \sqrt{2} e^{3u}$. 

\subsubsection{Subalgebra $(\hat{Q}_1, \hat{Q}_3)$}
The solution of the quantum equations for this subalgebra is
\begin{align}
\Psi = e^{i v \kappa_1 + i \kappa_3 \phi} e^{-\frac{u}{2}} \left(
A J_{-\lambda} (6 e^u)  +  B J_{\lambda} (6 e^u)
 \right),
\end{align}
where $\lambda = \sqrt{3}\sqrt{-\kappa_1^2 -\kappa_3^2 }$. Since the wave function is not in the polar form, we consider approximations for the Bessel functions. The Bessel argument $e^{6 u}$ is small for small values of the time and diverges for large values of time. For small arguments, the form of the wave function becomes
\begin{equation}
\Psi_{sm} \approx c e^{-u/2} \cosh (\lambda u) e^{i (\kappa_1 v + \kappa_3 \phi)},
\end{equation}
while for large it takes the form
\begin{equation}
\Psi_{la} \approx c e^{-u} \cosh (6 e^u) e^{\kappa_1 v + \kappa_3 \phi}.
\end{equation}
The quantum potential in both cases is non-zero and equals to $\mathcal{Q}_{sm}  = \frac{1}{144} e^{-2u} (1+12 \kappa_1^2 +12 \kappa_3^2)$ for small arguments and $\mathcal{Q}_{la} = 1$ for large and we have assumed that $A_{13} = B_{13}$. Thus, the spacetime element is expected to differ from the classical in both cases. The phase function is the same for both of these limiting cases and equal to $S_{13} = \kappa_1 v + \kappa_3 \phi$. Solving the system of semiclassical equations with the above mentioned phase function we find that 
\begin{align}
u = c_1,\quad
\phi  = c_2 + \frac{\kappa_3 v}{\kappa_1},\quad
n  = \frac{6 e^{2 u} \dot{v}}{\kappa_1},
\end{align}
which still contain a gauge freedom due to the parametrization invariance of the model. In the gauge $n=e^{2u}$ we also find that 
\begin{equation}
v = \frac{\kappa_1 t}{6} + c_3,
\end{equation}
and using the inverse of the transformation \eqref{coord_trans_bianchi_V} and absorbing the redundant constants the final spacetime metric becomes
\begin{align}
ds^2 = -\frac{e^{3c_1}}{36} dT^2 + e^{c_1} dx^2 + e^{-2x + \frac{\kappa_1 T}{6}} dy^2 + e^{-2x - \frac{\kappa_1 T}{6}} dz^2 .
\end{align}
This metric contains two essential constants. The Ricci scalar is
\begin{equation}
R = \frac{1}{2 e^{c_1}} (\kappa_1^2 -12 e^{2 c_1}),
\end{equation}
while the covariant derivative of the corresponding Riemann tensor vanishes. Since all other scalars constructed from the Riemann tensor are also constant, we conclude that this metric does not have any singularities at all. 

\subsubsection{Subalgebra $\hat{Q}_2$}
In this case the Wheeler-DeWitt equation and the eigenvalue equation for the operator $\hat{Q}_2$ do not specify completely the wave function. In order to do this, we promote the Casimir invariant of the full algebra to operator as in the case of the $(\hat{Q}_2,\hat{Q}_4)$ subalgebra of the Bianchi type II model. The quantum Casimir operator is 
\begin{equation}
\hat{Q}_{13} = \frac{1}{2} (\hat{Q}_1^2 + \hat{Q}_3^2),
\end{equation}
which in this case does not coincide with the kinetic part of the Hamiltonian constraint as it happens in most of the cases we study. The corresponding eigenvalue equation is 
\begin{equation}
\hat{Q}_{13} \Psi  = \frac{1}{2} (- \partial_{\phi \phi} -\partial_{vv}) \Psi = \kappa_{13} \Psi,
\end{equation}
which is Hermitian under the same measure $\mu$ as the other operators. The result is 
\begin{equation}
\Psi= e^{i \kappa_2 \tan^{-1} (\frac{v}{\phi})} e^{\frac{u}{2}} \left( A_2 J_{-\lambda} (6 e^u) +B_2 J_{\lambda} (6 e^u)  \right) J_{\kappa_2} (\sqrt{2 \kappa_{13} (v^2 + \phi^2)}),
\end{equation}
where $\lambda= i \sqrt{6 \kappa_{13}}$. This solution has Bessel functions with different arguments. The argument $e^u$ has already been studied in the case of the two-dimensional subalgebra. The argument $u^2 + \phi^2$ has similar behaviour. Assuming that $A_2 = B_2$, we find that the wave function becomes 
\begin{align}
\Psi_{sm} \approx \exp \left(i \kappa_2 \tan^{-1} \left(\frac{v}{\phi}\right) -\frac{u}{2}\right)  \cos \left( \sqrt{6 \kappa_{13}} u \right) (v^2 + \phi^2)^{\frac{\kappa_2}{2}},
\end{align}
for small arguments, while it takes the form
\begin{align}
\Psi_{la} \approx \exp \left( i \kappa_2 \tan^{-1} \left(\frac{v}{\phi}\right) -\frac{u}{2}\right) \left( 
\frac{\cos e^u + \sin e^u}{(v^2 + \phi^2)^{1/4}} \sin \left(\sqrt{2 \kappa_{13} (v^2 + \phi^2)} -\frac{1}{2} \pi \kappa_2 + \frac{\pi}{4} \right)
\right),
\end{align}
for large. The phase function is the same for the two cases 
\begin{align}
S = \kappa_2 \tan^{-1} \left(\frac{v}{\phi}\right).
\end{align}
The quantum potential does not vanish for both cases, thus indicating that the semiclassical solution will not be the same as the classical. Indeed, we find 
\begin{align}
u& = c_1,\\
v &= \sqrt{2 c_2 -\phi^2} ,\\
\phi & = \frac{\sqrt{2 c_2} }{\sqrt{1+ \tan^2 ( \frac{12 c_2 c_3 - \kappa_2 t}{12 c_2} )}} \tan \left( \frac{12 c_2 c_3 - \kappa_2 t}{12 c_2} \right).
\end{align}
By making the choice of gauge $n = e^{2u}$ we acquire the following spacetime element
\begin{align}
ds^2 = - \frac{c_2 e^{3 c_1}}{\kappa_2^2 T (2 c_2 -T)} dT^2 + e^{c_1} dx^2 + e^{-2 x + \sqrt{T} + c_1} dy^2 + e^{-2 x - \sqrt{T} + c_1} dz^2 ,
\end{align}
which has three essential constants (which are not the same as the ones in the classical case). The Ricci scalar is 
\begin{equation}
R= - \frac{\kappa_2^2 T + 2 c_2 \kappa_2^2 + 48 c_2^2 e^{2 c_1}}{8 c_2^2 e^{3 c_1}} .
\end{equation}
There is no singularity at $T \rightarrow \infty$ because we demand $T< c_2$ in order for the metric to remain of Lorentzian signature, but there is one for $T \rightarrow -\infty$.  Thus time never reaches infinite values where the spacetime would encounter a singularity. 

\section{Bianchi type VI coupled to a massless scalar field}\label{bianchi_VI}
\subsection{Classical analysis}
The Bianchi type VI class A model has the general form of spacetime metric of eq. \eqref{bianchi_metric} with invariant one-forms being
\begin{equation}
\sigma =\left(
\begin{array}{ccc}
 0 & e^{-x} & 0 \\
 0 & 0 & e^{x} \\
 1 & 0 & 0 \\
\end{array}
\right),
\end{equation}
while the $\gamma$ matrix has the diagonal form $\gamma = diag (a^2,b^2,c^2)$. The 4-dimensional spacetime is therefore
\begin{equation}\label{bianchi_vi}
ds^2 = -N^2 dt^2 + b^2 dx^2 + e^{-2 x} a^4 dy^2 + e^{-2 x} b^4 dz^2 .
\end{equation}
The axisymmetric form of \eqref{bianchi_vi} for this Bianchi type is given by taking into account the solution of the Einstein equation $G^1_2=T^1_2$,  which leads to the condition $c=b$. The integrand of the total action \eqref{total_action}, after discarding a term of total derivative, becomes 
\begin{equation}
L = -\frac{2 a^2 N}{b} -\frac{2 b \dot{a}^2}{N} -\frac{4 a \dot{a} \dot{b}}{N} + \frac{a^2 b \dot{\phi}^2}{N} .
\end{equation}
In order adopt to the constant potential parametrization, we make the transformation $N=\frac{b n}{2 a^2}$, where $n$ is the lapse function in the new parametrization. The Lagrangian therefore becomes
\begin{equation}
L = -n -\frac{4 a^2 \dot{a}^2}{n} -\frac{8 a^3 \dot{a}\dot{b}}{b n} + \frac{2 a^4 \dot{\phi}^2}{n}.
\end{equation}
The corresponding Hamiltonian constraint is
\begin{align}
\mathcal{H}& = 1-\frac{b}{8a^3} p_a p_b + \frac{b^2}{16 a^4} p_b^2 + \frac{1}{8 a^4} p_\phi^2 \approx 0 .
\end{align}
and the supermetric
\begin{align}
G_{\alpha \beta} &= \left(
\begin{array}{ccc}
-8a^2 & -\frac{8 a^3}{b} & 0 \\
-\frac{8 a^3}{b} & 0 &0 \\
0 & 0& 4 a^4 \\
\end{array}
\right) .
\end{align}
This superspace is conformally flat and has the following Killing fields and homothecy
\begin{align}
\xi_1 &= a \partial_a - b \ln (a^2 b^4) \partial_b -2 \phi \partial_\phi,\quad
\xi_2 = b \phi \partial_b + 2 \ln a \partial_\phi, \\
\xi_3 &= b \partial_b,\quad
\xi_4  = \partial_\phi,\quad
\xi_h  = b (\ln (\sqrt{a} b)) \partial_b + \frac{\phi}{2} \partial_\phi .
\end{align}
These generators satisfy the Lie bracket algebra 
\begin{align}\label{poisson_bianchi_vi}
&[\xi_1,\xi_2]=2 (\xi_2 + \xi_4),\quad
[\xi_1,\xi_3]= 4 \xi_3, \quad
[\xi_1,\xi_4]= 2 \xi_4,\quad
[\xi_2,\xi_4]= - \xi_3,\\
&[\xi_1,\xi_h]=\frac{1}{2} \xi_3,\quad
[\xi_2,\xi_h]= \frac{1}{2} \xi_2,\quad
[\xi_3,\xi_h]= \xi_3, \quad
[\xi_4,\xi_h]= \frac{1}{2} \xi_4 .
\end{align}
The first integrals of motion in the velocity phase space are
\begin{align}
Q_1 &= \frac{1}{n} \left( (- 8 a^3 + 16 a^3 \ln a + 32 a^3 \ln b) \dot{a} - \frac{8 a^4}{b} \dot{b} - 8 a^4 \phi \dot{\phi} \right),\\
Q_2 & = -\frac{8 a^3 \phi}{n} \dot{a} + \frac{8 a^4 \ln a }{n} \dot{\phi},\\
Q_3 & = -\frac{8 a^3 \dot{a}}{n},\\
Q_4 &= \frac{4 a^4 }{n} \dot{\phi}, \\
Q_h & =- \frac{4 a^3 \ln (a b^2)}{n} \dot{a} + \frac{2 a^4 \phi}{n} \dot{\phi} - \int n \ dt .
\end{align}
The solution of the system $Q_i = \kappa_i, \ i=1,2,3,4,h$ is
\begin{align}
N &= \frac{b n}{2 a^3}, \\
a & =  \frac{1}{2^{1/4}} (8 c_1 -\kappa_3 t)^{1/4}, \\
\phi & = c_2 -\frac{\kappa_4}{2 \kappa_3} \ln (8 c_1 -\kappa_3 t),\\
b& = 2^{1/8} \exp \left(\frac{-2 \kappa_1 \kappa_3 + \kappa_3^2 - 8 c_1 c_3 \kappa_3^5- 4 c_2 \kappa_3 \kappa_4 + 2 \kappa_4^2 + 16 c_1 c_3 \kappa_3^2 \kappa_4^2 + c_3 \kappa_3^4 (\kappa_3^2 - 2 \kappa_4^2) t }{8 \kappa_3^2}\right) (8 c_1 -\kappa_3 t)^{-\frac{1}{8} + \frac{\kappa_4^2}{4 \kappa_3^2}},\\
c_3 &=- \frac{8}{\kappa_3^3 (\kappa_3^2 - 2 \kappa_4^2)},
\end{align}
where the gauge choice $n=1$ has been adopted. The last relation of constants is used to remove $c_3$ and the rest of them are absorbed by performing suitable coordinate transformations. The final line element is written
\begin{equation}
ds^2 = - \frac{e^{2T} T^{-\frac{5}{4} + \alpha}}{\beta} dT^2 + \frac{e^{2T} T^{\frac{1}{4} - \alpha}}{ 2 \beta} dx^2 + e^{-2x} \sqrt{T} dy^2 +  e^{2x} \sqrt{T} dz^2 ,
\end{equation}
where $\alpha, \beta$ are combinations of the previously existing constants. The Ricci scalar is 
\begin{equation}
R =- \frac{\alpha}{\beta} e^{-2T} T^{-\frac{3}{4}- \alpha} ,
\end{equation}
and the singularity appears for $\alpha >- \frac{3}{4}$ and $T \rightarrow 0$. 
\subsection{Quantization and semiclassical analysis}
The quantization of the system is performed in the new coordinates as in the previous cases. The admissible subalgebras are all the one-dimensional and the two-dimensional $(\hat{Q}_2, \hat{Q}_3)$ and $(\hat{Q}_3, \hat{Q}_4)$. We study the latter two and $\hat{Q}_1$ from the one-dimensional ones. The system of the quantum equations to be solved are
\begin{align}
\hat{Q}_1 \Psi &= - i \left( -2 \phi \partial_\phi - b \ln (a^2 b^4) \partial_b + a \partial_a \right) \Psi = \kappa_1 \Psi,\\
\hat{Q}_2 \Psi & = - i (\ln (a^2) \partial_\phi + b \phi \partial_b) \Psi = \kappa_2 \Psi,\\
\hat{Q}_3 \Psi & = - i b \partial_b \Psi = \kappa_3 \Psi,\\
\hat{Q}_4 \Psi & = - i \partial_\phi \Psi = \kappa_4 \Psi,\\
\hat{\mathcal{H}} \Psi & = \frac{1}{16 a^4} \left(16 a^4 -2 \partial_{\phi,\phi} + b (\partial_b -b \partial_{b,b} + 2 a \partial_{a,b}) 
\right) \Psi =0,
\end{align}
where the measure is $\mu = \frac{16 a^5}{b}$.

\subsubsection{Subalgebras $(\hat{Q}_2, \hat{Q}_3)$ and $(\hat{Q}_3, \hat{Q}_4)$}
The wave function for the first subalgebra is 
\begin{align}
\Psi = \frac{c_1}{a (\ln a)^{1/2}} \exp \left(i \left( \kappa_3 \ln (\sqrt{a} b) -\frac{(\kappa_2 -\kappa_3 \phi)^2 -8 a^4 \ln a}{4 \kappa_3 \ln a}  \right)\right),
\end{align}
while for the latter
\begin{align}
\Psi = \frac{c_1}{a} \exp \left( i (\kappa_3 \ln b + \frac{2 a^4}{\kappa_3} + \frac{\kappa_3}{2} \ln a + \frac{\kappa_4^2 \ln a}{\kappa_3})\right).
\end{align}
Even though the coefficient is not constant in both cases, $\Omega = \frac{c_1}{a (\ln a)^{1/2}}$ for the first case and $\Omega =\frac{c_1}{a} $ for the latter one, the quantum potential vanishes for both subalgebras, thus we obtain the classical solution. 
\subsubsection{Subalgebra $\hat{Q}_1$}
In this case, the solution is
\begin{align}
\nonumber \Psi&= a^{\frac{-1+ i \kappa_1}{2}} (-1 + \ln (a^4 b^8))^{\frac{-1- i \kappa_1}{8}} \\
&\Big( 
A_3 J_{-\lambda} (a^2 \sqrt{-1 + \ln (a^4 b^8)}) + A_4 J_{\lambda} (a^2 \sqrt{-1 + \ln (a^4 b^8)}) \\
\nonumber &+ a^{\frac{1+ i \kappa_1}{2}} (-1 + \ln (a^4 b^8))^{\frac{1+ i \kappa_1}{8}} \left( A_1 \cosh ( 2 \sqrt{2} a^2 \phi) +A_2 \sinh ( 2 \sqrt{2} a^2 \phi) \right) 
\Big) .
\end{align}
For small arguments the wave function becomes 
\begin{align}
\Psi_{sm} \approx (A_4 + A_1 \cosh (2 \sqrt{2} a^2 \phi) + A_2 \sinh (2 \sqrt{2} a^2 \phi)) e^{i \kappa_1 \ln a} + A_3 a^{-1} (-1 + \ln (a^4 b^8))^{-\frac{1}{4}} e^{i \frac{\kappa_1}{4} \ln (-1 + \ln (a^4 b^8))} .
\end{align}
The demand that the wave function does not take infinite values for small arguments leads to the conditions $A_3=0$, thus simplifying the phase function to $S_{sm} = \kappa_1 \ln a$. Similar considerations for the large arguments lead to the form of the wave function
\begin{align}
\Psi_{la} \approx (A_1 \cosh (2 \sqrt{2} a^2 \phi) + A_2 \sinh (2 \sqrt{2} a^2 \phi)) e^{ i \kappa_1 \ln a}.
\end{align}
It is thus evident that the phase function is the same for both approximations and the semiclassical analysis is common for both limits. The solution of the semiclassical equations is
\begin{align}
a =c_2, \quad b = \frac{c_3}{2 c_2^2} e^{- \frac{kappa_1 t}{8 c_2^4}}, \quad \phi=c_1
\end{align}
The spacetime metric after gauge fixing the lapse function $n=1$ and absorbing the inessential constants is
\begin{align}
ds^2 = -\frac{4 e^T}{\kappa^2} dT^2 + e^T dx^2 + e^{-2 x} dy^2 + e^{2x} dz^2 ,
\end{align}
where $\kappa = \frac{\kappa_1}{c_2^2}$. This spacetime has one essential constant and a singularity at $T \rightarrow - \infty$ as can be seen by the Riemann scalar
\begin{equation}
R = -2 e^{-T}.
\end{equation}

\section{Matter content of the semiclassical spacetimes}\label{Physical Interpretation}
\subsection{Introduction}
The line elements that emerge from the semi-classical analysis do not obey the Einstein's field equations coupled with the scalar field unless of course in the cases where the quantum potential $\mathcal{Q}$ is zero. In order to assign a possible physical meaning to the solutions obtained, we calculate the Einstein tensor $G_{ij}=R_{ij}-\frac{1}{2}R g_{ij}$ for the semiclassical metrics and interpret it as an effective energy-momentum tensor by writing $G_{ij}=T^{(imf)}_{ij}$. We then check if this $T_{ij}$ fits to that of an imperfect fluid; for the calculational details see \cite{1985Madsen,Madsen:1988ph,Pimentel:1989bm}. The energy-momentum of an imperfect fluid is
\bal
T^{(imf)}_{ij}=\left( \rho+p \right)u_iu_j+p g_{ij}+2q_{(i}u_{j)}+\pi_{ij},
\eal
where $\rho$ is the energy density of the fluid, $u_i$ the 4--velocity, $q_i$ the heat flux vector, $p$
the pressure and $\pi_{ij}$ the anisotropic stress tensor. The relations that make the identification possible are
\bsub
\begin{alignat}{2}
\Pi_{mn}&=G_{ij} h^i{}_m h^j{}_n=p\, h_{mn}+\pi_{mn} , 
&\quad\quad \pi_{mn}&=\Pi_{mn}-\frac{1}{3}\Pi_k{}^k h_{mn}=\Pi_{mn}-p h_{mn},\\
\rho&=G_{ij}u^i u^j,  & p&=\frac{1}{3}\Pi_i{}^i,\\
q_k&=-G_{ij}u^i h^j{}_k,
\end{alignat}
\esub
 where $h_{ij}$ is the projection tensor orthogonal to velocity $u_i$ defined by
\bal
h_{ij} = g_{ij}+ u_i u_j \quad \text{with} \quad u_i u^i=-1.
\eal
Furthermore the kinematical quantities of the fluid that are of interest and appear in the decomposition of the covariant derivative of the velocity \cite{ellis2012relativistic}
\bal
\nabla_i u_j=-\dot{u}_i u_j+\omega_{ij}+\sigma_{ij}+\frac{1}{3}\theta h_{ij},
\eal
are
\bal
\dot{u}_i=u^j\nabla_j u_i, \quad \theta=\nabla_i u^i, \quad \sigma_{ij}=\nabla_{(i} u_{j)}+\dot{u}_{(i}u_{j)}-\frac{1}{3}\theta h_{ij}, \quad \omega_{ij}=\nabla_{[i} u_{j]}+\dot{u}_{[i}u_{j]},
\eal
i.e. the acceleration, the expansion, the shear and the rotation of the fluid respectively.
\subsection{Kantowski-Sachs spacetime}
Writing down the Einstein equations with Ricci scalar \eqref{ricci_kantowski}, we find that the effective energy-momentum tensor for this spacetime has the following components: The anisotropic stress tensor is
\begin{align}
\pi_{ij} &= \left(
\begin{array}{cccc}
0 & 0 & 0 & 0 \\
0 & -\frac{2}{3 t^2} & 0 & 0 \\
0 & 0 & \frac{1}{3} & 0 \\
0 & 0 & 0 & \frac{\sin^2 \theta}{3} \\
\end{array}
\right),
\end{align}
while the heat flow $q_i =0$, thus the stress tensor does not mimic a perfect fluid. The pressure and energy density are respectively given by
\begin{align}
p=-\frac{3 + 4 \alpha}{12 \alpha \ t}, \qquad
\rho = \frac{4 \alpha -1}{4 \alpha \ tr},
\end{align}
and lead to an equation of state with constant parameter $w= \frac{3 + 4 \alpha}{3 -12 \alpha}$. Moreover the fluid has zero acceleration, exhibits no rotation and is not shear free, since the shear tensor $\sigma_{ij}$ reads
\bal
\sigma_{ij}=
\begin{pmatrix}
0 		& 0 								& 0 									& 0											\\
0		& \frac{2}{3\sqrt{\alpha t^3}}		& 0									& 0											\\
0		& 0								& -\frac{1}{3}\sqrt{\frac{t}{\alpha}}	& 0											\\
0		& 0								& 0									& -\frac{1}{3}\sqrt{\frac{t}{\alpha}}\sin^2\theta \\
\end{pmatrix}.
\eal

\subsection{Closed and open FLRW spacetime}
The effective energy-momentum tensor for this spacetime is that of a perfect fluid because the anisotropic stress tensor $\pi_{ij}$ as well as the heat flow $q_i$ vanishes. The pressure and energy density equal
\begin{align}
p=-k, \quad
\rho =3k,
\end{align}
thus giving an equation of state with constant parameter $w= -\frac{1}{3}$. Moreover the fluid has zero acceleration and expansion, exhibits no rotation and is shear free.

\subsection{Spatially flat FLRW spacetime}
Contrary to the results for the $k =\pm 1$ cases, the anisotropic stress tensor $\pi_{ij}$ for the spatially flat case does not vanish,
\begin{align}
\pi_{ij} &= \left(
\begin{array}{cccc}
0 & 0 & 0 & 0 \\
0 & -\frac{2}{3 r^2} & 0 & 0 \\
0 & 0 & \frac{1}{3} & 0 \\
0 & 0 & 0 &\frac{\sin^2 \theta}{3} \\
\end{array}
\right),
\end{align}
but the heat flow does so, $q_i =0$. The pressure and energy density in this case are respectively
\begin{align}
p=-\frac{1}{3 r^2}, \quad
\rho =\frac{1}{r^2},
\end{align}
thus the equation of state has a constant parameter $w= -\frac{1}{3}$. The stress tensor thus does not mimic a perfect fluid. Likewise as in  $k =\pm 1$ cases the fluid has zero acceleration and expansion, exhibits no rotation and is shear free.
\subsection{Bianchi type II spacetime}
In this case there are three different subalgebras which give a semiclassical spacetime different from the classical one. For the subalgebra $(\hat{Q}_1,\hat{Q}_6)$, the anisotropic stress tensor $\pi_{ij}$ does not vanishes and its components are functions of $e^t, t, x$ and powers of $t$ up to third order. The same also holds for the pressure and the energy density which finally give an equation of state with parameter 
\begin{equation}\label{state_param}
w= \frac{-\lambda_1 e^t -2 \lambda_2 t^3 + \lambda_2 (\lambda_2 -3 \lambda_3 ) t^2 + \lambda_3 (\lambda_2 - \lambda_5) t}{3 e^t \lambda_1 -3 \lambda_2 t (\lambda_2 t + \lambda_3)}.
\end{equation} 
Obviously the matter content is not a perfect fluid. 

For the $(\hat{Q}_1,\hat{Q}_6)$, the situation about the effective energy-momentum tensor is similar as in the previous algebra, with the functional form of the components of the anisotropic stress tensor as well as the pressure and the energy density being even more complicated functions of the coordinates thus giving a correspondingly complicated function for the parameter $w$. 

Finally, for the subalgebra $(\hat{Q}_2,\hat{Q}_5)$ and $({\hat{Q}_3, \hat{Q}_4})$, the form of these expressions are complicated but the parameter $w$ of the equation of state is similar to the \eqref{state_param}. Furthermore in all of the above cases the fluid has zero acceleration exhibits no rotation but the shear and the expansion take nonzero values.

\subsection{Bianchi type V spacetime}
For the subalgebra $\{ \hat{Q}_1, \hat{Q}_3\}$, the effective energy-momentum tensor mimics a perfect fluid since the anisotropic stress tensor $\pi_{ij}$ and the heat flow $q_i$ vanishes. The pressure and energy density are respectively 
\begin{align}
p=\frac{\kappa^2 -1}{4 \lambda}, \quad 
\rho =-\frac{3 \kappa^2 +1}{4 \lambda^2},
\end{align}
thus giving an equation of state with constant parameter $w= \frac{1-\kappa^2}{1+ 3 \kappa^2}$. The fluid has zero acceleration and expansion, exhibits no rotation is not shear free since the shear tensor assumes the form
\bal
\sigma_{ij}=
\begin{pmatrix}
0	& 0		& 0									&0 									\\
0	& 0		& 0									&0 									\\
0	& 0		&-\frac{1}{2}\lambda e^{	t-\kappa\,x}	&0 									\\
0	& 0		&0									&\frac{1}{2}\lambda e^{-t-\kappa\,x} 	\\
\end{pmatrix}.
\eal

For the spacetime emerging from the one-dimensional subalgebra $\hat{Q}_2$, the effective anisotropic stress tensor is 
\begin{align}
\pi_{ij} &= \left(
\begin{array}{cccc}
0 & 0 & 0 & 0 \\
0 & 0 & 0 & 0 \\
0 & 0 & \frac{\lambda}{8} e^{\sqrt{t} -2 x} & 0 \\
0 & 0 & 0 &-\frac{\lambda}{8} e^{-\sqrt{t} -2 x} \\
\end{array}
\right),
\end{align}
while the heat flow vanishes. Thus, the energy-momentum tensor does not mimic a perfect fluid. The pressure and energy density are
\begin{align}
p=-\frac{\lambda t + \kappa-16 }{16 \mu^2}, \quad
\rho = -\frac{\lambda t + \kappa +48}{16 \mu^2},
\end{align}
and the equation of state is one with a non constant parameter
\begin{equation}
w= \frac{\lambda t + \kappa-16 }{3 -12 \alpha}.
\end{equation} The fluid has zero acceleration and expansion, exhibits no rotation and this time the shear reads
\bal
\sigma_{ij}=
\begin{pmatrix}
0	& 0		& 0														&0 														\\
0	& 0		& 0														&0 														\\
0	& 0		&-\frac{1}{4}\mu \sqrt{\lambda t+\kappa} e^{\sqrt{t}-2\,x}	&0 														\\
0	& 0		&0														&\frac{1}{4}\mu \sqrt{\lambda t+\kappa} e^{-\sqrt{t}-2\,x} 	\\
\end{pmatrix}.
\eal

\subsection{Bianchi type VI spacetime}
For the algebra $\hat{Q}_1$, the effective energy momentum tensor does not mimic a perfect fluid since the resulting anisotropic stress tensor is
\begin{align}
\pi_{ij} &= \left(
\begin{array}{cccc}
0 & 0 & 0 & 0 \\
0 & -\frac{4}{3} & 0 & 0 \\
0 & 0 & \frac{2}{3} e^{-t -2 x} & 0 \\
0 & 0 & 0 &\frac{2}{3} e^{-t + 2 x} \\
\end{array}
\right).
\end{align}
The pressure and energy density are
\begin{align}
p&=-\frac{e^{-t} }{3}, \\
\rho &= -e^{-t},
\end{align}
indicating an equation of state with constant parameter $w= - \frac{1}{3}$. Finally the fluid has zero acceleration, exhibits no rotation has expansion $\theta=-\kappa e^{-t/2}/4$ and shear tensor
\bal
\sigma_{ij}=
\begin{pmatrix}
0	& 0								& 0								&0 								\\
0	& -\frac{1}{6}\kappa e^{t/2}		& 0								&0 								\\
0	& 0								&\frac{1}{12}\kappa e^{-t/2-2x}	&0 								\\
0	& 0								& 0								&\frac{1}{12}\kappa e^{-t/2+2x} 	\\
\end{pmatrix}.
\eal

\section{Discussion}\label{conclusions}
In this paper we implemented the canonical quantization procedure, via the use of conditional symmetries, for various cosmological models minimally coupled to a massless scalar field. Particularly, we investigated the classical and quantum solutions in the cases of a (flat/non flat) FLRW space-time, Kantowski-Sachs and three axisymmetric Bianchi models (II, V and VI).

The methodology followed: we first construct the mini-superspace description of the models, by integrating the dynamically irrelevant degrees of freedom out of the initial action. At this point, the classical integrals of motion of the constraint system are obtained i.e. quantities that are conserved modulo the quadratic constraint of the system. As it is seen in the cases under consideration, these symmetries (when enough) allow for the derivation of the classical solution, avoiding the second order differential equations of motion. We proceed with the quantization by employing the canonical quantization procedure and following Dirac's prescription for systems with constraints. In addition, we promote to operators the generators of the existing classical symmetries, and apply them as supplementary conditions together with the Wheeler-DeWitt equation. The subsets of the symmetry generators that can be used in this way are dictated by the integrability condition \eqref{selection_rule} which renders the ensuing system of first order partial differential equations consistent. After the system of the eigenvalue equations plus the quadratic constraint is solved, we proceed with a semiclassical analysis for each case. This is based on Bohm's method for acquiring semiclassical trajectories, by using the phase of the wave function as an effective action for the system under consideration. The interesting outcome of the semiclassical analysis is that we do not always end up with the classical solution, thus we can interpret the resulting spacetime as one that contains a  different matter content from the standard classical, i.e. a massless scalar field. We have chosen to interpret the matter content as that corresponding to an imperfect fluid, which differs from the perfect one by the addition of a heat flux vector $q_i$ along with an anisotropic stress tensor $\pi_{ij}$. The result was that in all cases the heat flux was zero while there were cases in which the anisotropic stress tensor was different from zero, i.e. in Kantowski--Sachs model, in spatially flat FLRW model, in Bianchi type $VI$, $II$ and in one of type $V$ models. The interesting point is that in all but the two last models the equation of state of the fluid is described by a constant parameter $w$. Finally there were the closed and open FLRW models along with one of Bianchi type $V$ model that behave like \emph{perfect fluids} with a constant parameter $w$, a quite unexpected but certainly welcomed result, since in the classical level we started with a massless scalar field and in the quantum level we end up with a perfect fluid.

A reason behind this work has been the investigation of whether the existence of spacetime singularities can be suppressed at the semiclassical level; thus indicating that the quantum description can indeed remedy this problem of classical cosmology. A few notes are pertinent here. First, for most of the systems taken under consideration, there exists a subalgebra (the larger admissible) of the symmetry generators that gives rise to the exact classical spacetime through this procedure. Hence, for these specific subalgebras, no quantum corrections take place, at least at the level of the particular analysis: a fact that it can be considered as both good - since it shows that we are in accordance with the classical theory - and bad, in the sense that it does not rectify the classical problem. Nevertheless, in all cases, there exist alternative choices of admissible subalgebras for which the emerging spacetime is not identified with the classical solution. It is in these particular representations that we may expect to find a modified behavior of the semiclassical system concerning curvature singularities. Second, for the latter subalgebras there has been followed a different strategy on selecting the gauge. In some of our models the gauge choice was driven by the demand that the lapse function of the final 4-dimensional spacetime be the same as in the classical line element while in the rest of the models this demand has not been imposed.

In most of the models, all singularities seem to be successfully screened out by the quantization procedure. In the FLRW case we acquired either a flat spacetime (spatially flat case) or one that its higher derivative curvature scalars are zero, while the Riemann curvature scalars are constant. Therefore, in this case, the semiclassical spacetimes are non singular. The same also holds true for the Bianchi V model. However, in the other examples the situation is not so perfect. In the Bianchi VI model, the singularity can only be avoided by a proper choice of the essential constants. We can also see that the same happens in the Kantowski - Sachs as well as the Bianchi II case. 

It is noteworthy that, as we demonstrate in the appendix \ref{comparisonbarvinsky}, the wave functions we find following this procedure coincide with the results obtained in \cite{PhysRevD.89.043526}. In particular, the wave functions \eqref{wave_mom} coincide with equations (71) and (118) in the cases $\bold{(a),(b)}$ respectively. It is interesting that these results come from different view points. In both cases a scaling of the supermetric was performed; however in \cite{PhysRevD.89.043526} the wave function \eqref{wave_mom} emerges from a Wheeler-DeWitt equation with a manifest flat supermetric while here the flatness emerges after the rescaling of the supermetric in order to accomplish a constant potential.

Another interesting point for discussion is the choice to perform the Bohmian analysis by considering the wave function as a pure eigenstate of the operators $\hat{Q}_i$. This treatment has been employed partly because it is usually considered that the outcome of a measurement/observation is the eigenvalue of the physical quantity in question and partly because it keeps the calculations at a fairly manageable level. Indeed, the most general solution is a superposition of the eigenstates. In the cases the operator imposed has a continuous spectrum, this has several calculational as well as conceptual problems. The most straightforward way to solve these problems would be to construct wave packets and ``read off" the principal function $S$ from the final form it will take. This could constitute the subject of a future work. To take a quick look ahead, we can comment that: in the case of the spatially flat FLRW, the superposition of two eigenstates results in a principal function of the form $S \approx (\lambda_1 + \lambda_2) \phi$, where $\lambda_i, \ i=1,2$ are two different values of the constant $\kappa_3$. This form is similar form to $S \approx \kappa_3 \phi$ obtained when considering only one eigenstate. 

In summary, we can claim that the method applied here seems to have in general a positive effect in the extinction of singular points from the classical metric. Nevertheless, we have to keep in mind that this is largely an approximate treatment. After all, we substitute a gravitation theory with a dynamically equivalent mechanical system and quantize the latter. However, the ``regularization" appearing in semiclassical space-times produced by these quantum mechanical systems is a good indication of what should be expected by the full quantization of a gravitational theory.

\section*{Acknowledments}
The authors have the pleasure to thank Dr. N. Dimakis for fruitful discussions and crucial observations on a previous form of the paper. 

\appendix
\section{Limiting values of the Bessel functions}
\begin{enumerate}[(i)]
\item Bessel function of the first kind $J_\nu (z)$ is defined as
\begin{equation}
J_\nu (z) = \left(\frac{z}{2}\right)^\nu \sum_{k=0}^{\infty} (-1)^k \frac{(\frac{z^2}{4})^k}{k! \Gamma(\nu + k +1)}
\end{equation}

for $z \rightarrow 0$
\begin{equation}
J_\nu (z) \approx \frac{1}{\Gamma (\nu +1)} \left(\frac{z}{2}\right)^\nu,\quad 0< z << \sqrt{\nu +1}
\end{equation}

for $z \rightarrow \infty$
\begin{align}
J_\nu (z) &\approx \sqrt{\frac{2}{\pi z}} \cos (z- \frac{\nu \pi}{2} - \frac{\pi}{4})
\end{align}

\item Bessel function of the second kind Y
\begin{equation}
Y_\nu (z) =\frac{J_\nu (z) \cos (\nu z) - J_{-\nu} (z)}{ \sin (\nu \pi) }
\end{equation}

for $z \rightarrow 0$
\begin{equation}
Y_\nu (z) \approx \frac{1}{\Gamma (\nu +1)} \left(\frac{z}{2}\right)^\nu,\quad 0< z << \sqrt{\nu +1}
\end{equation}

for $z \rightarrow \infty$
\begin{align}
Y_\nu (z) &\approx \sqrt{\frac{2}{\pi z}} \sin (z- \frac{\nu \pi}{2} - \frac{\pi}{4})
\end{align}

\item modified Bessel function $I_\nu (z)$ defined by the relation
\begin{equation}
I_\nu (z) = \left(\frac{z}{2}\right)^\nu \sum_{k=0}^{\infty} \frac{(\frac{z^2}{4})^k}{k! \Gamma(\nu + k +1)}
\end{equation}

for $z \rightarrow 0$
\begin{equation}
I_\nu (z) \approx \frac{1}{\Gamma (\nu +1)} \left(\frac{z}{2}\right)^\nu,\quad 0< z << \sqrt{\nu +1}
\end{equation}

for $z \rightarrow \infty$
\begin{align}
I_\nu (z) &\approx \frac{e^z}{\sqrt{2 \pi z}}
\end{align}

\end{enumerate}

\section{Wave functions for the spatially flat FLRW case with constant potential}\label{comparisonbarvinsky}

In \cite{PhysRevD.89.043526}, Barvinsky \& Kamenshchik examined three different FLRW models\footnote{With our choice of the signature of the metric, i.e. $(-+++)$.}
\bal\label{FLRWBar}
ds^2=-N^2(t)dt^2+e^{2\alpha(t)}d\Omega_3,
\eal
coupled with a massless scalar field $\phi(t)$ in the presence of a constant potential $V(\phi)$ with the aid of the action
\bal\label{BarAct}
S=\int\! d^4x\sqrt{-g}\left( \frac{R}{\kappa}-\frac{\beta}{2} g^{\mu\nu}\phi_{,\mu}\phi_{,\nu}+\epsilon V_0 \right),
\eal
where $\kappa=1/(16\pi G)$ and the choices $\bold{(a)}$ $(\beta=1,\epsilon=-1)$, $\bold{(b)}$ $(\beta=1,\epsilon=0)$ and $\bold{(c)}$ $(\beta=-1,\epsilon=1)$ correspond respectively to the cases of a scalar field with negative potential, a scalar field with a vanishing potential and a ghost field with a positive potential. The supermetric for these models admits three Killing fields $\xi_{(I)}$ along with a homothetic field $\xi_h$ which read
\bal\label{homBar}
&\xi_1 = e^{-3\alpha+\frac{1}{2}\sqrt{3\beta}\phi}\partial_a - \frac{2\sqrt{3}}{\sqrt{\beta}}e^{-3\alpha+\frac{1}{2}\sqrt{3\beta}\phi} \partial_\phi,\non \quad 
\xi_2 = e^{-3\alpha-\frac{1}{2}\sqrt{3\beta}\phi}\partial_a + \frac{2\sqrt{3}}{\sqrt{\beta}}e^{-3\alpha-\frac{1}{2}\sqrt{3\beta}\phi} \partial_\phi, \\
&\xi_3 = \partial_\phi, \quad 
\xi_h = \frac{a}{6}\partial_a
\eal
which in turn they yield the integrals of motion $Q_{(I)}=\xi_{(I)}^\mu p_\mu$. The quantum analogues of $Q_{(I)}$ span a Lie algebra with abelian subalgebras $(\hat{Q}_1),(\hat{Q}_2),(\hat{Q}_3),(\hat{Q}_1,\hat{Q}_2)$. All but the $\hat{Q}_3$ of these subalgebras lead to wave functions with zero quantum potential $\mathcal{Q}$, thus leaving as the only nontrivial case the wave function resulting from the subalgebra $\hat{Q}_3$. For the cases $(a),(c)$\footnote{The case $(b)$ is treated in section \ref{Massless field in the closed FLRW universe}.}, the wave function has the form
\bal\label{waveBar}
\Psi_{\kappa_3}(\alpha,\phi)= e^{i\,\kappa_3\,\phi}\left(\psi_1(\kappa_3)\,J_{-s}\left(e^{3\alpha} \sqrt{\frac{8\epsilon\,V_0}{3\kappa}}\right) +\psi_2(\kappa_3)\, J_s\left( e^{3\alpha}\sqrt{\frac{8\epsilon\,V_0}{3\kappa}}    \right) \right)\,\text{where} \,  s=\frac{2i\,\kappa_3}{\sqrt{3\beta}}.
\eal
The general solution for the wave function should be a superposition of the above eigenstates, i.e
\bal\label{full_wave_fun}
\Psi(\alpha,\phi)=\int_{-\infty}^{+\infty} \! d\kappa_1\Psi_{\kappa_1}(\alpha,\phi)=\int_{-\infty}^{+\infty} \! d\kappa_1 e^{i\,\kappa_1\,\phi}f_{\kappa_1}(\alpha).
\eal
In order to find the wave function $\Psi(\alpha,p_\phi)$ in the momentum representation we take the Fourier transform of \eqref{full_wave_fun}
\bal\label{wave_mom}
\Psi(\alpha,p_\phi) &=\frac{1}{2\pi}\int_{-\infty}^{+\infty}\!d\phi \int_{-\infty}^{+\infty}\! d\kappa_1 e^{-ip_\phi \phi}\Psi_{\kappa_1}(\alpha,\phi)\non\\
                           & = \frac{1}{2\pi}\int_{-\infty}^{+\infty}\!d\phi \int_{-\infty}^{+\infty}\! d\kappa_1 e^{-ip_\phi \phi}e^{i\,\kappa_1\,\phi}f_{\kappa_1}(\alpha)\non\\
                           & =\int_{-\infty}^{+\infty}\! d\kappa_1 f_{\kappa_1}(\alpha) \delta(p_\phi-\kappa_1) \Rightarrow \non\\
\Psi(\alpha,p_\phi) &=\psi_1(p_\phi)\,J_{-s}\left(e^{3\alpha} \sqrt{\frac{8\epsilon\,V_0}{3\kappa}}\right) +\psi_2(p_\phi)\, J_s\left( e^{3\alpha}\sqrt{\frac{8\epsilon\,V_0}{3\kappa}}    \right)  \,\text{where} \,  s=\frac{2i\, p_\phi}{\sqrt{3\beta}}.
\eal
It is worth investigating the fact that this approach seems to select the non trivial wave functions (i.e. those wih non vanishing quantum potential) found with the approach of the present work. 
\bibliography{axisymmetric}
\end{document}